\documentclass[sigconf]{acmart}

\usepackage{pifont}
\usepackage{microtype}
\usepackage{graphicx}
\usepackage{subfigure}
\usepackage{booktabs} 
\usepackage{paralist}
\usepackage{amsmath}
\usepackage{empheq}
\usepackage{mdframed}
\usepackage[utf8]{inputenc}
\usepackage[english]{babel}
\usepackage{amsthm} 
\usepackage{xspace}
\usepackage{bm}
\usepackage{multirow}

\usepackage{algorithm}
\usepackage[noend]{algorithmic}

\usepackage{color, colortbl}
\usepackage{cellspace}
\newtheorem{problem}{Problem}

\newcommand\scalemath[2]{\scalebox{#1}{\mbox{\ensuremath{\displaystyle #2}}}}

\theoremstyle{definition}

\newcommand{\hide}[1]{}

\makeatletter
\newcommand\footnoteref[1]{\protected@xdef\@thefnmark{\ref{#1}}\@footnotemark}
\makeatother

\renewcommand{\algorithmicrequire}{\textbf{Input:}}
\renewcommand{\algorithmicensure}{\textbf{Output:}}
\renewcommand{\algorithmiccomment}[1]{\hfill$\blacktriangleright$ #1}

\newcommand{\cbit}{\begin{compactitem}}
	\newcommand{\ceit}{\end{compactitem}}
\newcommand{\cben}{\begin{compactenum}}
	\newcommand{\ceen}{\end{compactenum}}

\newcommand{\beq}{\begin{equation}}
	\newcommand{\eeq}{\end{equation}}

\newcommand{\beqn}{\begin{equation*}}
	\newcommand{\eeqn}{\end{equation*}}

\newcommand{\bit}{\begin{itemize}}
	\newcommand{\eit}{\end{itemize}}
\newcommand{\ben}{\begin{enumerate}}
	\newcommand{\een}{\end{enumerate}}

\newcounter{x}

\newcommand{\map}{{\tt map}\xspace}

\newcommand{\df}{{\tt DF}\xspace}

\newcommand{\open}{{\it OSM}\xspace}
\newcommand{\gis}{{\it Gisette}\xspace}
\newcommand{\surl}{{\it SpamURL}\xspace}

\newcommand{\mF}{\mathcal{F}}
\newcommand{\mFt}{\mathcal{F}^{(t)}}
\newcommand{\mFtm}{\mathcal{F}^{(t-1)}}

\newcommand{\mH}{\mathcal{H}}
\newcommand{\mP}{\mathcal{P}}

\newcommand{\mC}{\mathcal{C}}

\newcommand{\bdel}{\boldsymbol{\Delta}}

\newcommand{\bfl}{\mathbf{f}}
\newcommand{\bx}{\mathbf{x}}

\newcommand{\br}{\mathbf{r}}
\newcommand{\bs}{\mathbf{s}}
\newcommand{\bzb}{\bar{\mathbf{z}}}
\newcommand{\bz}{\mathbf{z}}

\newcommand{\be}{\bm{\epsilon}}

\newcommand{\R}{\mathbb{R}}

\newcommand{\Z}{\mathbb{Z}}

\newcommand{\method}{{\sc Spar}x\xspace}
\newcommand{\ex}{{\sc xStream}\xspace}

\newcommand{\dbs}{{\sc DBSCOUT}\xspace}
\newcommand{\ddlof}{{\sc DDLOF}\xspace}
\newcommand{\dif}{{\sc SPIF}\xspace}

\newcommand{\conmod}{{\tt config-mod}\xspace}
\newcommand{\congen}{{\tt config-gen}\xspace}

\newcommand{\eps}{{\tt eps}\xspace}
\newcommand{\minpts}{{\tt minPts}\xspace}

\newcommand{\la}[1]{{\color{blue}#1}}

\AtBeginDocument{%
	\providecommand\BibTeX{{%
			\normalfont B\kern-0.5em{\scshape i\kern-0.25em b}\kern-0.8em\TeX}}}

\copyrightyear{2022}
\acmYear{2022}
\setcopyright{rightsretained}
\acmConference[KDD '22] {Proceedings of the 28th ACM SIGKDD Conference on Knowledge Discovery and Data Mining}{August 14--18, 2022}{Washington, DC, USA.}
\acmBooktitle{Proceedings of the 28th ACM SIGKDD Conference on Knowledge Discovery and Data Mining (KDD '22), August 14--18, 2022, Washington, DC, USA}
\acmPrice{}
\acmISBN{978-1-4503-9385-0/22/08}
\acmDOI{10.1145/3534678.3539076}

\begin{document}

\title{\method: Distributed Outlier Detection at Scale}

\author{Sean Zhang}
\affiliation{%
  \institution{Carnegie Mellon University}
  \country{}
}
\email{xiaoronz@alumni.cmu.edu}

\author{Varun Ursekar}
\affiliation{%
	\institution{Carnegie Mellon University}
	\country{}
}
\email{vursekar@andrew.cmu.edu}

\author{Leman Akoglu}
\affiliation{%
	\institution{Carnegie Mellon University}
	\country{}
}
\email{lakoglu@andrew.cmu.edu}

\begin{abstract}
There is no shortage of outlier detection (OD) algorithms in the literature, yet a vast body of them are designed for a single machine. With the increasing reality of already cloud-resident datasets comes the need for distributed OD techniques.
This area, however, is not only understudied but also short of public-domain  implementations for practical use.
This paper aims to fill this gap: We design \method, a data-parallel OD algorithm suitable for shared-nothing infrastructures, which we specifically implement in Apache Spark. 
Through extensive experiments on three real-world datasets, with several billions of points and millions of features,
we show that existing open-source solutions fail to scale up; either by large number of points or high dimensionality, whereas 
\method yields scalable and effective performance.
To facilitate practical use of OD on modern-scale datasets, we open-source \method under the Apache license.\textsuperscript{\ref{note1}}
\end{abstract}



\maketitle

\section{Introduction}
\label{sec:intro}

{\bf Motivation.~}
Outlier detection (OD) finds applications in many domains, such as finance \cite{anandakrishnan2018anomaly}, manufacturing \cite{susto2017anomaly}, surveillance \cite{sodemann2012review} and environmental monitoring \cite{xu2020anomaly}, 
to name a few.
OD is typically used in data cleaning to filter out noise/errors/etc. before fitting a model that may be sensitive to the presence of outliers in the training data. In other settings, outliers are rather the ``signal'', where OD is employed to identify adversarial or abnormal occurrences, such as network attacks or about-to-fail manufacturing parts.

With the advent of technology, it is typical of applications to generate or collect 
massive datasets. Often these are already resident in modern distributed infrastructures or cloud services (e.g. Amazon AWS, Microsoft Azure, Google Cloud, etc.), which renders OD algorithms designed for a single machine inapplicable.  
These datasets may also be ever-growing, with new data being generated in a daily fashion or much faster; such as sensors monitoring data (including social sensors), transactions data, computer network logs data, among others.
This trend puts OD algorithms applicable to distributed large-scale datasets in great demand, which is likely to  grow further in coming years.


{\bf Prior Work.~}
While outlier detection has an extensive literature \cite{aggarwal2017introduction}, \textit{distributed} OD is a considerably understudied area.  
To our knowledge, there exists only a few published work on OD for cloud-resident data on shared-nothing infrastructures.
Among those, only a handful of them provide
public-domain implementations: DDLOF \cite{yan2017distributed}, a distributed implementation of the popular LOF \cite{breunig2000lof} OD algorithm; SPIF \cite{tao2018parallel} is a Spark-based design of the popular (ensemble) algorithm  Isolation Forest (IF) \cite{liu2008isolation}; and most recently, DBSCOUT \cite{corain2021dbscout} that builds on the ideas from the popular clustering algorithm DBSCAN \cite{ester1996density}.
DDLOF is based on Hadoop \cite{hadoop}, which is 10-100$\times$ slower than the in-memory computing platform Apache Spark \cite{Zaharia10}.
On the other hand, the Spark-based SPIF only
employs {\em model-parallelism} (training each ensemble component on a separate compute node); and because it does not leverage data-parallelism, it scales poorly to large datasets with numerous points. 
DBSCOUT is also built on Spark, however it {scales extremely poorly with increasing dimensionality $d$, and has been tested on 2- and 3-$d$ data only.} Moreover, {it is a distance-based algorithm that may not work well on data with varying-density support, 
	inherits two sensitive hyperparameters from DBSCAN, 
	and provides only a binary output (inlier/outlier) based on a strict outlier definition.}
(See Sec. \ref{sec:related} for detailed related work, and Sec. \ref{sec:experiments} for comparison experiments.)

Besides $i)$ scaling-out to \textit{distributed} datasets, there are several other desired characteristics of an OD algorithm for practical usability, including: 
$ii)$ \textit{linear} time and space complexity, 
$iii)$ \textit{robustness} to hyperparameter choices (so that practically easy to use in unsupervised settings), 
$iv)$ carefully handling \textit{high dimensionality}, and 
$v)$ admitting data with \textit{mixed-type features} (both categorical and numerical).
Today, no existing OD algorithm in the literature satisfies all these desired properties. 

{\bf Present Work.~}
Through this work, we set out to make OD a greater contributor to the real-world use cases at large, and
introduce a new OD algorithm called \method, \textit{exhibiting all the aforementioned practical properties i)--v)}.
Specifically, we capitalize on the \ex algorithm \cite{manzoor2018xstream} which is originally designed for a single machine, and transform it to a MapReduce \cite{dean2008mapreduce} based 
distributed algorithm. Our implementation is based on the Python API of Apache Spark (hence the name, \method) that is suitable for a shared-nothing distributed infrastructure.

\method readily inherits all the desirable properties of \ex, while also scaling-out to massive datasets on cloud platforms.
In fact, and apart from the extensive experiments in the original paper \cite{manzoor2018xstream}, \ex has recently been externally validated to outperform a long list of (9) baselines at router-level anomaly detection tasks on 64 different real-world datasets from Huawei Inc. \cite{navarro2021human}.
The algorithms are run using various hyperparameter (HP) configurations, and compared with respect to both best hyperparametrization (optimistic and unrealistic in unsupervised settings) as well as 
practical (just one) hyperparametrization (realistic yet conservative) as only one HP configuration (as recommended by the original authors) is used for all datasets. \ex performance is outstanding based on their extensive evaluation, quoting: ``Remarkably, \ex stands out for being close to the Ideal Ensemble Upper Bound [i.e. best model among all algorithms and HPs]'', and
``\ex [...] 
able to provide robust and good performance even in practical settings.''

We summarize the main contributions of this work as follows.

\cbit 
\item \textbf{Distributed OD for Cloud-resident Data}: We present \method, a \textit{data-parallel}
outlier detection (OD) algorithm for {distributed} 
data that 
handles large number of input points as well as high dimensionality.
\method is a  linear time and space complexity algorithm that can
scale-out to datasets that are already cloud-resident.
For wide-spread usability, we provide a public-domain implementation of \method in Apache Spark.\footnote{\label{note1}\la{\url{https://tinyurl.com/sparx2022}}}
\ceit

\cbit
\item \textbf{List of Desired Properties}: \method is effectively a distributed extension of the \ex algorithm which exhibits many desirable properties; including robustness to hyperparameters, handling high dimensional 
feature space, admitting mixed-type data, among others.
Arguably, all of those combined with scalability to cloud-resident datasets makes \method one of the most practically useful, open-source solutions to OD.

\item \textbf{Scalability and Effectiveness}: We evaluate \method against the state-of-the-art baselines 
DBSCOUT \cite{corain2021dbscout} (DBSCAN-centered OD in Spark) and
SPIF \cite{tao2018parallel} (Spark-based Isolation Forest).
As we show through 
extensive experiments on datasets with number of points and dimensionality up to several billions and millions,  
\method outperforms SOTA baselines in terms of detection accuracy, running time and  memory usage.
SPIF fails to scale up to large number of points (limited to model-parallelism), whereas
DBSCOUT scales poorly with high dimensionality (does not scale up beyond 10 dimensions). 
\ceit

\section{Preliminaries \& Background}
\label{sec:prelim}

Consider a distributed point-cloud database, originally containing $n$ points with $d$ dimensions; $\mP=\{\bx_1,\ldots,\bx_n\}$, where $\bx_i$'s can be {\em mixed-type}, i.e. a subset of features being real-valued and the rest being categorical with arbitrary domains. 
Let $\mF$ denote the set of original features where $|\mF|=d$.

We consider a general 
deployment 
setting where new features may arise over time; e.g. a new attack-indicator starts being tracked at time $t$, where $\mFt$ depicts the feature space at time $t$.
Specifically, at any time $t$,  
(1) new points $\bx_{n+1}, \bx_{n+2}, \ldots$ may arrive with dimensionality $|\mFt|\geq |\mFtm|$, as well as (2) points seen thus far may receive value-updates to arbitrary (including new) features; where $<ID, F, \delta>$ denotes an update-triple for point with identifier $ID$ to feature $F \in \mFt$ of value $\delta$.
For a real-valued feature $F$, $\delta \in \R$ is a value-update, whereas $\delta =\;$\texttt{\small{old\_val:new\_val}} is a value-substitution for a categorical feature (\texttt{\small{old\_val}} is \texttt{\small{null}} if $F$ is a newly-arising feature). For example, \texttt{\small{$<$id,URL,+3$>$}} may indicate a social media user with identifier \texttt{\small{id}} sharing \texttt{\small{3}} more posts containing a link to a certain \texttt{\small{URL}}.
Similarly,  \texttt{\small{$<$id,loc,NYC:Austin$>$}} may depict a customer with \texttt{\small{id}} relocating (substituting \texttt{\small{loc}})  from \texttt{\small{NYC}} to \texttt{\small{Austin}}. 

Vast majority of work for deployed OD systems assumes row-streams where newcoming points (i.e. rows) to be outlier-scored exhibit 
\textit{fixed} dimensionality. To distinguish our deployed setting, where not only new rows but also new columns (i.e. features) may arrive, we refer to such incoming data as {\em evolving streams}.

\vspace{-0.1in}
\subsection{Problem Statement}

We aim to tackle the OD problem for very large datasets that are stored in a distributed fashion on commodity machines.
This is typical of numerous settings where the data is prohibitively large to be stored on a single server or is already collected in a decentralized fashion (e.g. distinct customer bases around the globe).

We consider a \textit{shared-nothing} parallel computing environment on which the data is to be processed, typical of many modern big data infrastructures including Apache Hadoop and Spark \cite{hadoop,Zaharia10}. 
Here each
compute node (or worker machine) only has access to partial data, and processes it locally and independently of others(i.e. is idempotent), where intermediate results/data are then exchanged between the workers over the network.  
Computation typically alternates between several iterations of parallel local computation (e.g. a \texttt{map} phase) and communication (e.g. a \texttt{reduce} phase).
Parallelism and local computation help with fast processing, whereas network speed is low, hence,
network communication costs are often the bottleneck for distributed computing.
Common strategies to reduce network costs include reducing the number of iterations (e.g. by computing locally more, and communicating less frequently) and/or reducing the size of the intermediate results to be communicated.

As we will show in Sec. \ref{sec:dod}, \method is only a \textit{two-pass} algorithm, i.e. it requires only two iterations of \texttt{map} and \texttt{reduce} phases. Moreover, the intermediate data objects being passed between worker machines is of \textit{constant size}, further reducing the burden (i.e. time-delay) of network communication.

We build \method for massive-scale cloud-resident data, although by design it can also handle distributed evolving streams. We present static and streaming problem definitions separately below.  
In Sec. \ref{sec:dod} we describe the distributed algorithms underlying \method in detail, and discuss how to build on this solution to address streaming input when deployed in Sec. \ref{ssec:sd}. 

\vspace{-0.05in}
\begin{problem}[Distributed OD for Static Data]
	\textbf{Given} a  static point-cloud database $\mP=\{\bx_1,\ldots,\bx_n\}$ that 
	is stored in a distributed file system (very large $n$);
	\textbf{Design} a distributed OD algorithm to compute outlier scores $\{s_1,\ldots,s_n\}$ on a shared-nothing computing infrastructure, with \underline{linear} time and space complexity.
\end{problem}
\vspace{-0.1in}
\begin{problem}[OD on Incoming Data Streams]
	\textbf{Given} an 
	incoming stream at 
	$t=1,2,\ldots$, where (1) points with new $ID$ may arrive with 
	$|\mFt|\geq |\mFtm|$ of mixed-type features, or (2)
	point-wise $\delta$-updates 
	$<$$ID, F, \delta$$>$ may arrive to existing points with $ID$ and $F\in \mFt$; 
	\textbf{Compute} (updated) outlier score for $ID$, in \underline{constant} time.
\end{problem}

Before we delve into distributed algorithms, we provide a summary of the single-machine \ex OD algorithm in this section.

\vspace{-0.05in}
\subsection{\ex for Single-Machine OD: Summary}
\label{ssec:xstream}

\ex \cite{manzoor2018xstream} is designed for outlier detection in high-dimensional data streams.
In a nutshell, it consists of three main phases. First, to tackle high-dimensionality, it creates efficient data sketches, which can be computed on-the-fly even for newcoming features.
Then, it builds efficient counting data structures for histogram-based density estimation in random subspaces of the feature space. Finally, it performs
outlier scoring based on the approximated density estimates.
We provide details on each of these steps as follows.

\vspace{-0.05in}
\subsubsection{\bf Step 1. Data Projection}
\label{sssec:s1}

Let us begin by assuming the feature space (i.e. dimensionality) is fixed, such that $\bx_i \in \R^D$ where $D>d$ is the new dimensionality after one-hot-encoding (OHE) the categorical features. 
A low-dimensional sketch (or embedding) $\bs_i$ can be created for each point (while accurately preserving pairwise distances between the points) by random projections \cite{indyk1998approximate,achlioptas2003database}: 
\beq
\label{eq:sketch}
\bs_i = (\bx_i^T \br_1, \dots, \bx_i^T\br_K)
\eeq
where $\{\br_1,\ldots,\br_K\}$
depict $K$ random Gaussian vectors \cite{indyk1998approximate} or \textit{sparse} random vectors where with probability 1/3, $\br_k[F]\in \{\pm1\}$ and zero otherwise \cite{achlioptas2003database}.  
The latter choice not only is ``database-friendly'', i.e. more efficient to store and compute, but can be also advantageous for outlier detection by effectively looking at data subspaces, reducing the masking effect of irrelevant features \cite{zimek2012survey}.

Notice that the same $\br_k \in \R^D$'s are used for all points over a stream, and hence need to be cashed.
However, for evolving streams wherein new features may emerge, $D$ is not fixed and in fact unknown apriori.
Then, the idea is not to cash, but to hash. Specifically, entries of each $\br_k$ is to be computed on-the-fly via hashing, such that Eq. \eqref{eq:sketch} is rewritten as follows.
\beq
\label{eq:hash}
\scalemath{0.85}{
	\bs_i[k] =  \sum_{F \in \mF_r} h_k(F) \;\cdot\; \bx_i[F]  
	\;+ \sum_{F \in \mF_c}  h_k(F \oplus \bx_i[F]) \;\cdot\; 1
	\;,\;\; k=1\ldots K
}
\eeq
\vspace{-0.1in}

\noindent
where $h_k(\cdot)$ is a hash function, $\mF_r$ and $\mF_c$ denote the set of real-valued and categorical features, respectively, $\bx_i[F]$ is point $i$'s value of feature $F$, and $\oplus$ denotes the string-concatenation operator. 
Each hash function takes as input a string and returns $+1, -1$ or $0$ with respective probabilities $1/6, 1/6$ and 2/3.
(See \cite{manzoor2018xstream} for implementation details of such hash families.) 
For numerical features the input string is the feature name. For categorical features, it is the concatenation of the name and the corresponding feature value.
Effectively, the sparse random vector entries are computed for any feature via hashing, i.e. $\br_k[F]=h_k[F]$, and multiplied with the corresponding feature value. For categorical features, the concatenated string corresponds to the OHE feature name with value $1$.

When triplet updates $<ID, F, \delta>$ arrive over the stream, where $\delta =\;$\texttt{\small{old\_val:new\_val}} for categorical features, the sketch can be updated by
\begin{equation}
	\label{eq:update}
	\hspace{-0.125in}
	\scalemath{0.9}{
		\bs_{ID}[k] = \begin{cases}
			\bs_{ID}[k] + h_k(F) \cdot \delta   {\text{\quad if real-valued $F$,}}
			\\
			\bs_{ID}[k] - h_k(F\oplus\text{\texttt{\small{old\_val}}}) + h_k(F\oplus\text{\texttt{\small{new\_val}}})   \text{  o.w.} 
		\end{cases}
	}
\end{equation}
for $k=1\ldots K$ such that $h_k(F\oplus\text{\texttt{\small{old\_val}}})$ returns zero when \texttt{\small{old\_val}} is \texttt{\small{null}}.
It is important to notice that Eq. \eqref{eq:update} can seamlessly handle a newly emerging feature $F$ that has never been seen before.

\subsubsection{\bf Step 2. Half-space Chains}
\label{sssec:s2}

Anomaly detection relies on density estimation at multiple scales via a set of so-called Half-space Chains (HC), a data structure akin to multi-granular subspace histograms.
Each HC has a length $L$ (or $L$ layers), along which the (projected) feature space $\mF_p$ is recursively halved on a randomly sampled (with replacement) feature, where $f_l \in \{1,\ldots,K\}$ denotes the feature at level $l=1,\ldots, L$.
As such, a point can reside in one of 2 bins at level 1, one of 4 bins at level 2, and in general one of $2^l$ bins at level $l$.
Given the sketch $\bs$ of a point, the goal is to efficiently identify the bin it falls into at each level.

Let $\bdel \in \R^K$ be the vector of initial bin widths, which is equal to half
the range of the projected data along each dimension $f \in \mF_p$.
Let $\bzb_l \in \Z^K$ denote the bin identifier of $\bs$ at level $l$, initialized to all zeros.
At level 1, the bin-id is updated as $\bzb_1[f_1] = \lfloor \bs[f_1]/\bdel[f_1] \rfloor$.
In general, the bin-id at consecutive levels can be computed \textit{incrementally}, following

\vspace{-0.15in}
\begin{equation}
	\label{eq:binid}
	\scalemath{0.9}{
		\bzb_l[f_l] = \lfloor \bz_l[f_l] \rfloor \;\text{  s.t.  } \;
		\bz_l[f_l] = \begin{cases}
			\bs[f_l]/\bdel[f_l]  \text{\quad if $o(f_l,l) =1$, and}
			\\
			2\bz_l[f_l] \text{\quad \quad\;\; o.w.; if $o(f_l,l)> 1$ } 
		\end{cases}
	}
\end{equation}
where $o(f_l,l)$ denotes the number
of times feature $f_l = \{1,\ldots,K\}$ has been sampled in the chain until and
including level $l$.
We note that a small uniformly random value $\epsilon_l \in (0, \bdel[f_l])$, called shift, is added to the sketches at each level to remedy issues for nearby points around fixed bin boundaries. We omit those for brevity and refer to \cite{manzoor2018xstream} for details.

Notice that all points with the same unique $\bzb_l$ reside in the same histogram bin at level $l$. As such, level-wise (multi-scale) densities are to be estimated by counting the number of points with the same bin-id at each level. This can be done by a dictionary (or perfect hash) data structure.  
The number of possible bins, however, grows exponentally with $l$.
Even though data is not necessarily spread to all bins, number of non-empty bins (and hence the size of the dictionary) can grow very large for large $L$. Then, approximate counts can be obtained via a count-min-sketch \cite{cormode2005improved}, the size of which is user-specified, i.e. constant.

Overall, \ex is an ensemble of $M$ Half-Space Chains, $\mH = \{HC^{(m)} := (\bdel, \bfl^{(m)}, \be^{(m)},  \mC^{(m)})\}_{m=1}^M$ where each HC is associated with the following list of meta-data;
(i) the bin-width per feature $\bdel \in \R^K$,
(ii) the sampled feature per level $\bfl^{(m)} \in \Z^L$,
(iii) the random shift value per level $\be^{(m)} \in \R^L$, and (iv) the counting data structure per level $\mC^{(m)} = \{C_l^{(m)}\}_{l=1}^L$.

\subsubsection{\bf Step 3. Outlier Scoring}
\label{sssec:s3}
To score a given (updated) sketch for outlierness, the count of points in the bin that it falls into is identified at each level $l$ of a HC, denoted $C_l^{(HC)}[\bzb_l]$.
The count is extrapolated via multiplying by $2^l$ to estimate the total count if the data were distributed uniformly. Smallest estimate across levels\footnote{Note that the counts can be compared across levels after extrapolation.} is taken as the outlier score, and then averaged across all HCs as 

\vspace{-0.15in}
\beq
\label{eq:score}
\scalemath{0.9}{
	O(\bs) = \frac{1}{M} \sum_{m=1}^M \min_l \; 2^l \cdot C_l^{(m)}[\bzb_l] \;.
}
\eeq
A lower value indicates 
a granularity at which the point resides in a relatively sparse region, and hence 
higher outlierness.


\section{\method for Distributed OD}
\label{sec:dod}

In this section we introduce \method;
distributed algorithms corresponding to each of the 
three main steps (as described in Sec.s \S\ref{sssec:s1}--\S\ref{sssec:s3}) of \ex (Sec.s \S\ref{ssec:s1}--\S\ref{ssec:s3}),
space and time complexity analysis (\S\ref{ssec:st}), and 
OD on incoming data streams (\S\ref{ssec:sd}).

Our implementation uses Apache Spark's Python API. In Spark, the data points are stored in a distributed fashion across several compute nodes (or machines), residing in what-is-called a {\tt DataFrame} (\df).
The underlying distributed infrastructure is a shared-nothing architecture, consisting of worker nodes and a driver node, which can be implemented with a MapReduce programming paradigm.  The computation is broken down to pieces performed independently at worker nodes (typically through {\tt \textcolor{blue}{map}} operations), which may also exchange/communicate (over the network) intermediate data/results (typically through the {\tt \textcolor{red}{reduce}} operation).

\vspace{-0.1in}
\subsection{Step 1. Distributed Data Projection}
\label{ssec:s1}
The first step is to transform the input data to a $K$-dimensional representation through random projections. The projections are done based on Eq. \eqref{eq:hash} using random hash functions, $h_1(\cdot),\ldots, h_K(\cdot)$.  As projection of each point $i$ can be done independently using Eq. \eqref{eq:hash}, the workers can perform this step \textit{fully locally}, i.e. without any need for communication between workers. The projected data is stored in a new \df within the same worker nodes.

The pseudocode for Step 1 is given in Algorithm \ref{algo:step1}.
We first define a {\tt projector} with $K$ different seeds between $(0,1)$ and density 1/3 (Line 1).
Note that the same seeds are used across all the worker nodes to create new points in the same embedding space.  
Operationally, projection step involves a \textit{single} \map \textit{phase}. The \map operator takes as input a function and passes each element of the \df through it.  Namely the {\tt fit\_transform} function of the projector is fed to \map, which transforms an input point by projecting it $K$ times (Line 2). 
Steps of {\tt fit\_transform} is summarized in Lines 3--7, in which the {\tt hash\_string} function hashes the input string $str$ (i.e. the feature name) based on the input seed, and with probability 1/3 returns $\{\pm1\}$ and zero otherwise.

\renewcommand{\algorithmicrequire}{\textbf{Input:}}
\renewcommand{\algorithmicensure}{\textbf{Output:}}
\renewcommand{\algorithmiccomment}[1]{\hfill$\blacktriangleright$ #1}
\newcommand{\algrule}[1][.5pt]{\par\vskip.5\baselineskip\hrule height #1\par\vskip.5\baselineskip}
\makeatother

\begin{algorithm}[!t]
	
	\caption{\method \textbf{Step 1.} Distributed Data Projection }
	\label{algo:step1}
	\begin{algorithmic}[1]
		\small{
			\REQUIRE  inputDF (input data), $K$ (proj. dim.), featureNames
			\ENSURE projDF ($K$-dimensional transformed \df)
			\algrule
			\STATE projector = HashProjn($K$,  seeds=arange$(0, K, 1)$, density=1/3)	
			\STATE  projDF = inputDF.{\textcolor{blue}\map}(lambda $x$: projector.{\tt fit\_transform}($x$) 
			\algrule
			
			{\bf procedure} {\tt fit\_transform}($pt$)
			\STATE \hspace{0.1in} {{\bf for each} $F$ in featureNames:} 
			\STATE \hspace{0.2in} {{\bf if} $F$ is categoric} {\bf then }
			$F:= F \oplus pt[F]$  
			
			\STATE \hspace{0.1in} $R =$ array([[{\tt hash\_string}$(k, str)$ 
			for $k$ in seeds]
			\STATE \hspace{1.78in} for $str$ in featureNames])
			
			\STATE \hspace{0.1in} {\bf return}	 proj\_$pt$ =  $R \cdot pt$ 
			\COMMENT{dot product}	
		}			
	\end{algorithmic}
\end{algorithm}
\setlength{\textfloatsep}{0.05in}

\vspace{-0.1in}
\subsection{Step 2. Distributed Half-space Chains}
\label{ssec:s2}

After sparse projections we first obtain the feature ranges, specifically the gap between the minimum and maximum values in each of $K$ features of the projected \df, and set the bin-widths $\bdel \in \R^K$ to half of the ranges. The min and max values can be obtain easily for distributed data, by first finding those within each worker and then comparing the local min/max values across the workers.
Next we start creating the half-space chains, as shown in  Algorithm \ref{algo:step2}.

\vspace{-0.1in}
\subsubsection{\bf Data-parallel training of a single chain.~}
Given $\bdel$ and chain-length $L$ (Line 1),
each chain is instantiated at each level $l=1\dots L$ with a randomly picked split-feature $f_l$ from $\{1,\dots,K\}$ as well as a random shift amount $\epsilon_l \in (0, \bdel[f_l])$. These values are shared/common across the workers.
Then the workers start binning the points, by computing the unique $K$-dimensional bin-id of each point at every level.
Our implementation allows for constructing the histogram density estimation on a subsample of the data (Line 2).
As binning of points can be done independently, we implement this step through a \map operation (Line 3), which passes each point through the {\tt fit} function that returns the bin-id per level. The resulting bin-ids are stored in a new \df, called binIDsDF, in a distributed fashion.

Next we count the number of points that fall into each unique bin approximately, using a count-min-sketch (CMSketch) consisting of $r$ (numRows) hash-tables, each with $w$ (numCols) buckets (Line 4).
Each level of a chain is associated with a separate CMSketch (Line 5--6).
A CMSketch effectively hashes each input, i.e. a bin-id, into one of $w$ buckets.
Since collisions may occur, the hashing is repeated $r$ times using different hash functions.
The {\tt allCols} function of the CMS class in our implementation takes a bin-id as input, and  computes the bucket (i.e. column) index at each hash-table (i.e. row), returning an array of the form: 
\beq
\label{listpairs}
\hspace{-0.125in}[((1,colindx\_1),1),  ((2,colindx\_2),1),\dots, ((r,colindx\_w),1)]
\eeq
As each bin-id can be hashed independently, we operationalize binning through a map, more specifically {\tt flatMap}, which ``flattens'' the array of ({\tt key}=(row,col), {\tt value}=1) pairs returned by {\tt allCols} into individual elements in a \df (Line 7).
To sum up the count of points that hash into each bucket across workers, we perform a {\tt reduceByKey}, which groups all pairs with the same {\tt key} and sums their {\tt value}s (Line 8).
Finally, {\tt collectAsMap} gathers the total counts from across workers into the driver node (also Line 8).

\begin{algorithm}[!t]
	\caption{\method \textbf{Step 2.} Distributed Half-space Chains}
	\label{algo:step2}
	\begin{algorithmic}[1]
		\small{
			\REQUIRE projDF, $L$ (chain-length), sampleRate, numRows, numCols,  numChains, numThreads
			\ENSURE  CMSketches (counts at all levels $l=1\ldots L$ per chain) 
			\algrule
			{\bf procedure} {\tt fit\_chain}(seed)
			\STATE \hspace{0.1in}C = Chain($\bdel$, $L$)
			\STATE \hspace{0.1in}binIDsDF = projDF.rdd.sample(sampleRate, seed)
			\STATE \hspace{0.92in}.{\textcolor{blue}\map}(lambda $x$: C.{\tt fit}($x$))
			
			\STATE \hspace{0.1in}cms = CMS(numRows, numCols)
			\STATE \hspace{0.1in}{\bf for} $l$ in range($L$):
			\STATE \hspace{0.2in}C.CMSketches[$l$] = binIDsDF
			\STATE \hspace{0.4in}.{\textcolor{blue}{\tt flatMap}}(lambda $x$: cms.{\tt allCols}($x[l]$))
			\STATE \hspace{0.4in}.{\tt \textcolor{red}{reduceByKey}}(lambda a,b : a+b)
			.{\tt collectAsMap}()
			\algrule
			\STATE tpool = ThreadPool(numThreads)
			\STATE  indxlist = list(range(numChains))
			\STATE  tpool.{\textcolor{blue}\map}(lambda cind: {\tt fit\_chain}(cind), indxlist)
			
		}			
	\end{algorithmic}
\end{algorithm}
\setlength{\textfloatsep}{0.05in}

\vspace{-0.1in}
\subsubsection{\bf Model-parallel training of ensemble of chains.~} As described, training of each chain leverages \textit{data-parallelism} where we perform binning
on partitions of data in parallel across worker nodes. Only the intermediate results, in this case partial sums/counts of points per bin are pooled over the network from all workers.
\method is an \textit{ensemble} of such half-space chains, each of which can be trained independently. In a single-machine implementation, these chains are built in sequence within a for-loop. To foster further speed up, we add to our implementation \textit{model-parallelism} where the chains are trained by a pool of parallel threads (Lines 9--11). 



\vspace{-0.1in}
\subsection{Step 3. Distributed Outlier Scoring}
\label{ssec:s3}

Having collected all the partial counts across workers, the final CMSketches containing approximate bin counts reside in the driver node. 
For scoring, those are passed to individual workers which can then \textit{locally} compute the outlier score for each data point that they store.
That is, outlier scoring involves a single \map operation as outlined in Algorithm \ref{algo:step3}.

Being an ensemble, \method scores a point against each chain (Line 1).
The main computation occurs during \map, which passes each point through the {\tt score} function of the chain to which we also feed as input argument the respective CMSketch containing the (approx.) bin counts at all levels $l=1\dots L$ (Line 2).
Notably, we define CMSketches as a broadcast variable (Line 4), which tells Spark to pass it to workers \textit{only once} with each {\tt score} function call, since it is a fixed data structure that scoring does not alter.

For brevity, we do not include the steps for {\tt score} in the pseudocode, which we briefly describe here in text. The steps are very similar to those for {\tt fit} (i.e., identify the bin-id per level) and {\tt allCols} (i.e., identify the bucket that bin-id hashes to per hash table, at each level). 
Each bucket is associated with a total count (held in CMSketch of the chain), which is an overestimate for the count of points with the same bin-id due to collisions. Therefore, the \textit{minimum} count across the hash-tables is taken as the most accurate (i.e., least overestimate)---hence the name, count-\textit{min}-sketch.

To obtain a point's outlier score per chain,
the min-count at each level $l$ is extrapolated by $2^l$, and the smallest of the extrapolated counts is returned as the score from the current chain as in Eq. \eqref{eq:score}.

Similar to model-parallel fitting of the chains, we also score each point against the chains via a parallel thread pool (Lines 4--6). Differently, outlier score from each chain (i.e. thread) is summed across the pool (Line 6) and 
then averaged as in Eq. \eqref{eq:score}.

\begin{algorithm}[!t]
	\caption{\method \textbf{Step 3.} Distributed Outlier Scoring}
	\label{algo:step3}
	\begin{algorithmic}[1]
		\small{
			\REQUIRE projDF, CMSketches (for all chains)
			\ENSURE  outlier\_scores
			\algrule
			{\bf procedure} {\tt score\_chain}(cindex, CMS)
			\STATE \hspace{0.01in} C = Chains[cindex]
			\STATE \hspace{0.01in} scoreC=projDF.{\textcolor{blue}\map}(lambda $x$:C.{\tt score}($x$, CMS.value[cindex]))
			
			\algrule
			\STATE CMS = sc.broadcast(CMSketches)
			\STATE tpool = ThreadPool(numThreads)
			\STATE  idx = list(range(numChains))
			\STATE  outlier\_scores = {\tt sum}(tpool.{\textcolor{blue}\map}(lambda c: {\tt score\_chain}(c), idx))
		}			
	\end{algorithmic}
\end{algorithm}
\setlength{\textfloatsep}{0.05in}


\vspace{-0.1in}
\subsection{Space and Time Complexity}
\label{ssec:st}

We analyze space and time complexity for each step of \method, and present
storage and computation requirements both locally (per worker) and distributed (collectively across workers).

At the beginning, $n$ $d$-dimensional data points are stored in a Spark DataFrame, taking $O(nd)$ \underline{distributed}-storage. 
In \textbf{Step 1}, each feature is hashed $K$ times using different seeds\footnote{Note that it is enough to hash the name of numerical features only once, whereas categorical feature names are concatenated by the value a point takes (Algo. 1, Line 4) and hence are (re)hashed per point.}, where the resulting matrix $R$ (Algo. 1, Line 5) takes $O(Kd)$ \underline{local}-storage.
Projecting one point is a dot product, with $O(Kd)$ complexity. Thus, Step 1 takes $O(Kdn)$ \underline{distributed}-computation overall. 

In \textbf{Step 2}, all $M$ chains of the ensemble are trained in parallel by a thread pool, thus we multiply the all complexities below by $M$.
The workers first compute the $K$-dim. bin-id of each point at each of $L$ levels, hence the resulting binIDsDF takes $O(MKLn)$ \underline{distributed}-storage. A bin-id per level is computed incrementally using Eq. \eqref{eq:binid}, in $O(L)$ total time per point, and $O(MLn)$ \underline{distributed}-computation.
Next each point's bin-id is hashed $r$ times at each level by the {\tt allCols} function, for a total of $O(KrLMn)$ \underline{distributed}-computation.
The output is $r$ pairs as in \eqref{listpairs}
per point per level, taking $O(MrLn)$ \underline{distributed}-storage.
The {\tt reduceByKey} operation (Algo. 2, Line 8) groups all pairs with the same key=(row,col) in the same reducer node and sums the values, effectively finding the total count. This requires $O(MrLn)$ network communication, and $O(MrLn)$ \underline{distributed}-computation.

There are at most $r$(numRows)$\times$$w$(numCols) unique (row,col) keys which is equal to the user-specified CMSketch size. At the end,  {\tt collectAsMap} gathers these total counts across all layers and all chains at the driver node, requiring $O(rwLM)$ \underline{local}-storage.

In \textbf{Step 3}, final CMSketches are passed from the driver to each worker for scoring, requiring $O(rwLM)$ \underline{local}-storage. Scoring of a point has similar computational footprint to fitting where we, per level:   
create its bin-id, hash and read the counts from $r$ buckets it hashes to, take the minimum and extrapolation. Minimum extrapolated count across $L$ layers (i.e. outlier score) is averaged across $M$ chains. Overall it takes $O(KrLMn)$ \underline{distributed}-computation.

{\bf Remark:} Note that \method is not only linear in data size ($n$ and $d$) but also fully \textit{data-parallel}---all bigO terms involving $n$ are \underline{distributed}.
Moreover, all of $K, L, r, w, M$ are user-specified, thus space and time associated with those can be adjusted on demand. 

\vspace{-0.1in}
\subsection{OD on Incoming Data Streams}
\label{ssec:sd}

Upon deployment, a single compute node can serve as the front-end to receive and score newcoming point updates over an evolving stream. 
It requires $O(rwLM)$ {local}-storage to keep all CMSketches in-memory. For each $\delta$-update, the sketch-
and then the score-update takes $O(K)$ and $O(KrLM)$, respectively.
As existing points receive $\delta$-update, a size-$N$ LRU cache of IDs is maintained, along with their sketches for $O(NK)$ space. 
Note that both space and time complexity per update are constant, as all terms 
are user-specified.


\section{Experiments}
\label{sec:experiments}


\hide{
	We design experiments to evaluate \method with respect to comparison to SOTA baselines, speed-up by parallelism, and scalability.
	
	\cbit
	\item  How does \method compare to SOTA distributed OD baselines w.r.t. accuracy-vs-resources used, specifically  outlier detection rate, running time, and memory usage?
	
	\item  What is the speed-up achieved by \method with increasing parallelism?
	
	\item  How does \method scale with increasing data size (both $n$ and $d$) as compared to other distributed OD baselines?
	
	\ceit
}

\subsection{Setup}
\label{ssec:setup}

\subsubsection{\bf Datasets.~}

We experiment with three public-domain datasets,
varying largely in number of points $n$ and dimensions $d$, as well as in the fraction of outliers.
Summary statistics are given in Table \ref{tab:data}.

(1) \gis is a handwritten digits dataset, originally from the UCI ML repository, 
which has also been used for outlier detection.\footnote{\url{https://github.com/cmuxstream/cmuxstream-data}}
It has reasonably large number of features $d$, however its number of samples $n$ is very small (in fact, smaller than $d$), at least for testing distributed algorithms.
Following the outlier benchmark creation procedure by Steinbuss and B{\"o}hm \cite{steinbuss2021benchmarking}, we fit a Gaussian Mixture Model (GMM) to the 3500 inliers it originally contains. Then we draw $n$$=$40,000 samples from the fitted GMM such that around 10\% constitutes the outliers---inliers are drawn directly, while for generating outliers we increase the variance of 10\% of the randomly chosen features by a factor of 5 as recommended in \cite{steinbuss2021benchmarking}. This ensures that 90\% of the features do not convey any information on outlierness, making the detection task harder.
We use the resulting dataset as a small-$n$/large-$d$ testbed.

(2) \open \cite{haklay2008openstreetmap} is a 2-d dataset depicting the GPS coordinates (latitude, longitude) collected from OpenStreetMap (OSM) contributors\footnote{\url{https://blog.openstreetmap.org/2012/04/01/bulk-gps-point-data/}}.
With nearly 3 billion points and 51.5GB total size,  it is one of the largest real-world GPS datasets.\footnote{Data is downloaded from \url{https://planet.osm.org/gps/simple-gps-points-120312.txt.xz}}
We use \open as a very large-$n$/very small-$d$ dataset.
Besides its size, we use this low-dimensional dataset as it has been used for testing distributed OD algorithms previously \cite{yan2017distributed,corain2021dbscout}. As we will show, these approaches are limited to such low-$d$ settings and scale poorly with dimensionality. 

The original \open data does not contain any labeled outliers, and has been used in prior work \cite{yan2017distributed,corain2021dbscout} as is, only for measuring runtime and scalability.
We aim to study the complete landscape/trade-off between accuracy-versus-resources used. Therefore, to measure detection performance we inject into \open simulated outliers, as described in detail in Appx. \ref{appx:inject}. 
The visualization of the resulting dataset is shown in Fig. \ref{fig:osm}. 

(3) \surl \cite{conf/icml/MaSSV09} is a large-$n$/very large-$d$ dataset, which contains malicious and benign URLs along with numerous lexical and host-based characteristics of each URL as the features.
This dataset makes the detection task challenging not only computationally, owing to its size, but also statistically as outliers are likely buried in small subspaces of the high dimensionality.

\begin{table}[!t]
	\caption{Datasets used in experiments.}  
	\label{tab:data}
	\vspace{-0.15in}
	\centering
	\hspace{-0.1in}
	\scalebox{0.85}{
		\begin{tabular} {l|rrrcr} 
			\toprule
			\textbf{Name} &  $n$ \textbf{pts.} & $d$ \textbf{dim.} & \textbf{size (GB)} & \textbf{type} & \textbf{outl.} \\ 
			\midrule
			\gis & 40,000 & 4,971 & 4.69 & small-$n$/large-$d$ & 10\%\\
			\hline
			\open & 2,772,233,904 & 2 & 51.50 &  large-$n$/small-$d$ & 0.036\% \\
			
			\hline
			\surl & 2,396,130 & 3,231,962 & & large-$n$/large-$d$ & 33\% \\		
			\bottomrule
		\end{tabular}
	}
\end{table}

\vspace{-0.1in}
\subsubsection{\bf Baselines.~} 
Only a few OD algorithms are designed for distributed data and also have public-domain implementations.
We compare \method to two such SOTA distributed OD methods. 

(1) \dbs \cite{corain2021dbscout}: This is the most recent distributed OD algorithm, implemented in Spark using the Java API\footnote{\url{https://github.com/mattecora/dbscout}}. It uses  ideas  that are largely inspired by the popular DBSCAN clustering algorithm  \cite{ester1996density}, and constructs a cellular grid structure to parallelize and speed up the outlier identification process.

Another recent public-domain distributed OD approach, based on the popular LOF algorithm \cite{breunig2000lof},
is \ddlof \cite{yan2017distributed}. We omit it from our baselines for two reasons; first, \dbs has been shown to outperform \ddlof significantly on various datasets including our \open, and second its implementation is in Hadoop\footnote{\url{https://github.com/yizhouyan/DDLOFOptimized}} which is not on par with the efficiency of Apache Spark. 




(2) \dif \cite{tao2018parallel}: Today, Spark is notably more popular than Hadoop providing orders of magnitude speed-up. 
Isolation Forest (IF) \cite{liu2008isolation} is also one of the most popular OD algorithms owing to its competitive performance, as confirmed by evaluation studies \cite{emmott2015meta,ting2017defying}. 
Therefore, 
we also compare to \dif \cite{tao2018parallel}, a Spark-based design of IF, using a public-domain implementation.\footnote{\url{https://github.com/titicaca/spark-iforest}}

In a nutshell, IF subsamples data points to build a forest/ensemble of extremely randomized trees. In \dif, each tree is trained in parallel at a single worker using its respective subsampled data. The subsample for each tree is gathered at a single worker via a map-reduce phase, where $<$\texttt{\small{tree-ID}}, \texttt{\small{point}}> pairs are generated during \texttt{map}, and a \texttt{reduceByKey} is performed to shuffle all points needed to construct a tree to a single reducer/worker (!). As such, tree fitting is {\bf not} data-parallel; rather, forest construction is designed to be model-parallel. For large subsample sizes, this implementation quickly becomes infeasible as it shuffles too much intermediate data over the network (which is very slow) prior to model fitting.

\vspace{-0.1in}
\subsubsection{\bf Performance metrics.~} 
To measure performance, we study the accuracy-vs-resource requirements landscape of the algorithms with varying HP configurations, since these typically have a trade-off relation. For accuracy, we report the \textit{outlier ranking quality} using Area Under the ROC (AUROC), and Precision-Recall Curve (AUPRC), as well as F1 score. In terms of resources, we measure the \textit{running time} and \textit{peak memory usage}. 

\begin{figure}[!t]
	\centering
	\includegraphics[width=0.9\linewidth]{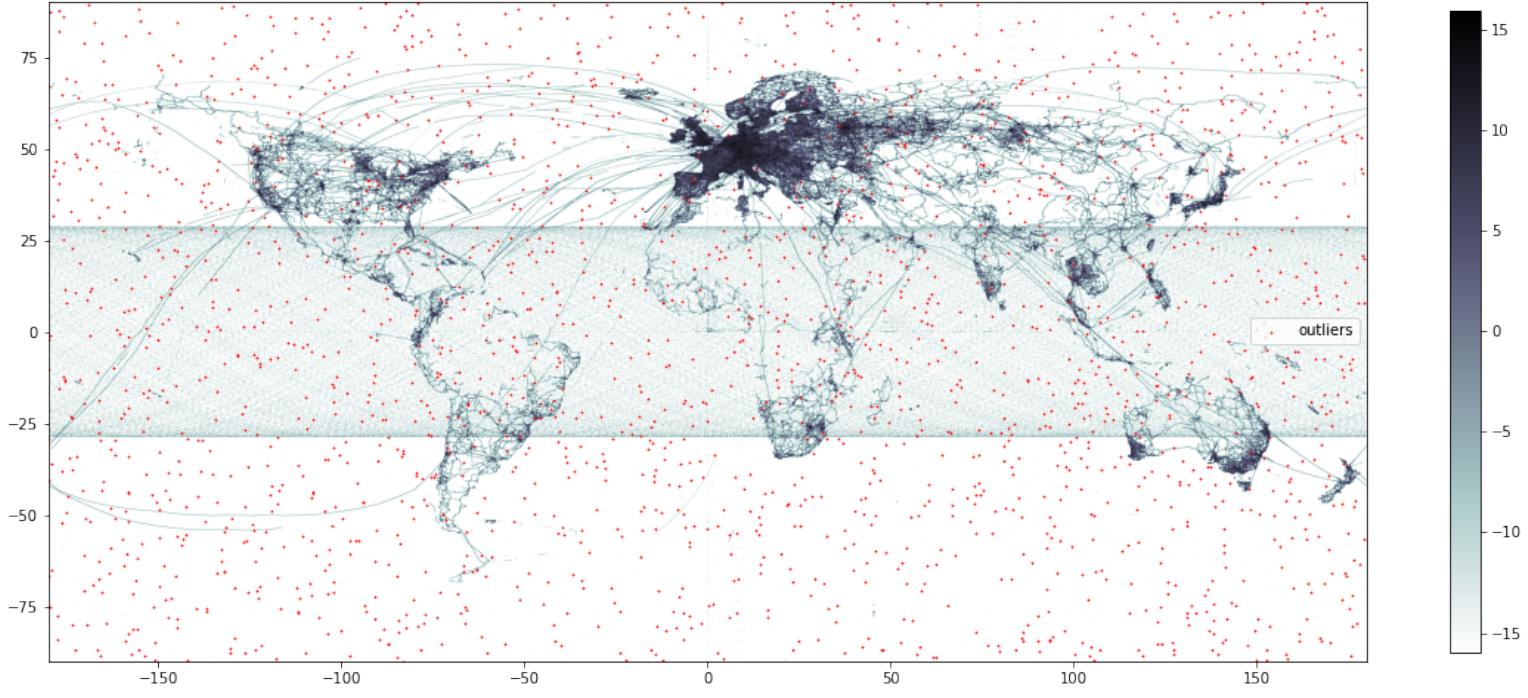}
	\vspace{-0.15in}
	\caption{Visualization of the 2-d \open dataset: (in black) inliers from real-world GPS traces, (in red) injected outliers. \label{fig:osm}}
	\vspace{-0.15in}
\end{figure}

\vspace{-0.1in}
\subsubsection{\bf System settings.~} 
We conduct experiments on the U.S. National Science Foundation Pittsburgh Supercomputing Center (PSC), and 
set up both a `moderate' (\conmod) as well as a `generous' system configuration (\congen) with relatively more resources, as specified in Table \ref{tab:sysconfigs} in Appx.

Specifically, we vary the number of data (i.e. \df) partitions, available memory for the driver as well as the worker (executor) machines, the number of executors and number of cores per executor, as well as the number of threads available for multi-threading. \congen has access to strictly more resources in all these aspects, typically doubling or more.

\vspace{-0.1in}
\subsubsection{\bf Model settings.~} 
We also run experiments with various hyperparameter (HP) configuration of the methods, as there is no clear means to setting those in unsupervised tasks.
Our proposed \method is similar to \dif in terms of being an ensemble of chains and trees, respectively, of certain depth, which can be trained on subsamples of data.
Accordingly we vary the number of ensemble components $M$ (tree or chain) (\{50,100\}), the depth or number of levels $L$ (\{10,20\}), and subsampling rate (\{0.01,0.1,1\}) for \method and \dif.
We use a fixed CMS size of $r$$=$$10$$\times$$w$$=$$100$ for \method on all datasets.
The number of projections is set to $K$$=$50 for \gis, and $K$$=${100} for \surl, while \open is not transformed, since it is already very low dimensional.

On the other hand, \dbs has two HPs; \eps and \minpts.
We vary \minpts 
and identify a corresponding \eps via the process explained in \cite{corain2021dbscout}, with quadratic (!), complexity: we plot the sorted distance to the \minpts-th neighbor across all points.  \eps is then chosen in the uppermost part of the ``elbow'' zone of the plot.

\subsection{Results}
\label{ssec:results}

\subsubsection{\bf \dbs does not scale w.r.t. dimensionality $d$.~} 
We start with analyzing results on our small-$n$/large-$d$ \gis. First, we show that \dbs scales very poorly with increasing dimensionality using this moderately large dimensional dataset. 

\begin{table}[!t]
	\caption{\dbs scales poorly with $d$. On \gis, running time grows fast from $d$$=$1 to 10, and then 
		times out.}  
	\label{tab:dbsdim}
	\vspace{-0.15in}
	\centering
	\scalebox{0.85}{
		\begin{tabular} {l|rr} 
			\toprule
			\textbf{$d$ dim.} &  \textbf{Runtime (sec)} & \textbf{Peak memory (MB)}  \\ 
			\midrule
			2                    & 11.3                        & 1,650                 \\
			4                    & 13.0                          & 1,630                 \\
			6                    & 31.1                        & 133,000               \\
			8                    & 429.8                       & 254,000               \\
			10                   & 3,420.0                        & 350,000               \\
			11                   & {TIMEOUT} & N/A \\
			\bottomrule
		\end{tabular}
	}
	\vspace{0.05in}
\end{table}

As shown in Table \ref{tab:dbsdim}, when using \congen, the running time grows dramatically fast as we run \dbs on \gis using an increasing number of randomly sampled 1--10 features. It reaches around 60-min mark at only $d$$=$10, and with 11 features, the process times-out (after 8 hrs). 
Peak memory usage responds similarly growing from 1.6GB to a total of 350GB across the executors.

Poor scalability makes \dbs infeasible for datasets with more than a handful of features, therefore we only report results on \gis for \method and \dif. 

\vspace{-0.05in}
\subsubsection{\bf Accuracy vs. Resources Landscape.~~} 
\begin{figure}[!ht]
	\vspace{-0.1in}
	\centering
	\hspace{-0.1in}	\includegraphics[width=0.4995\linewidth]{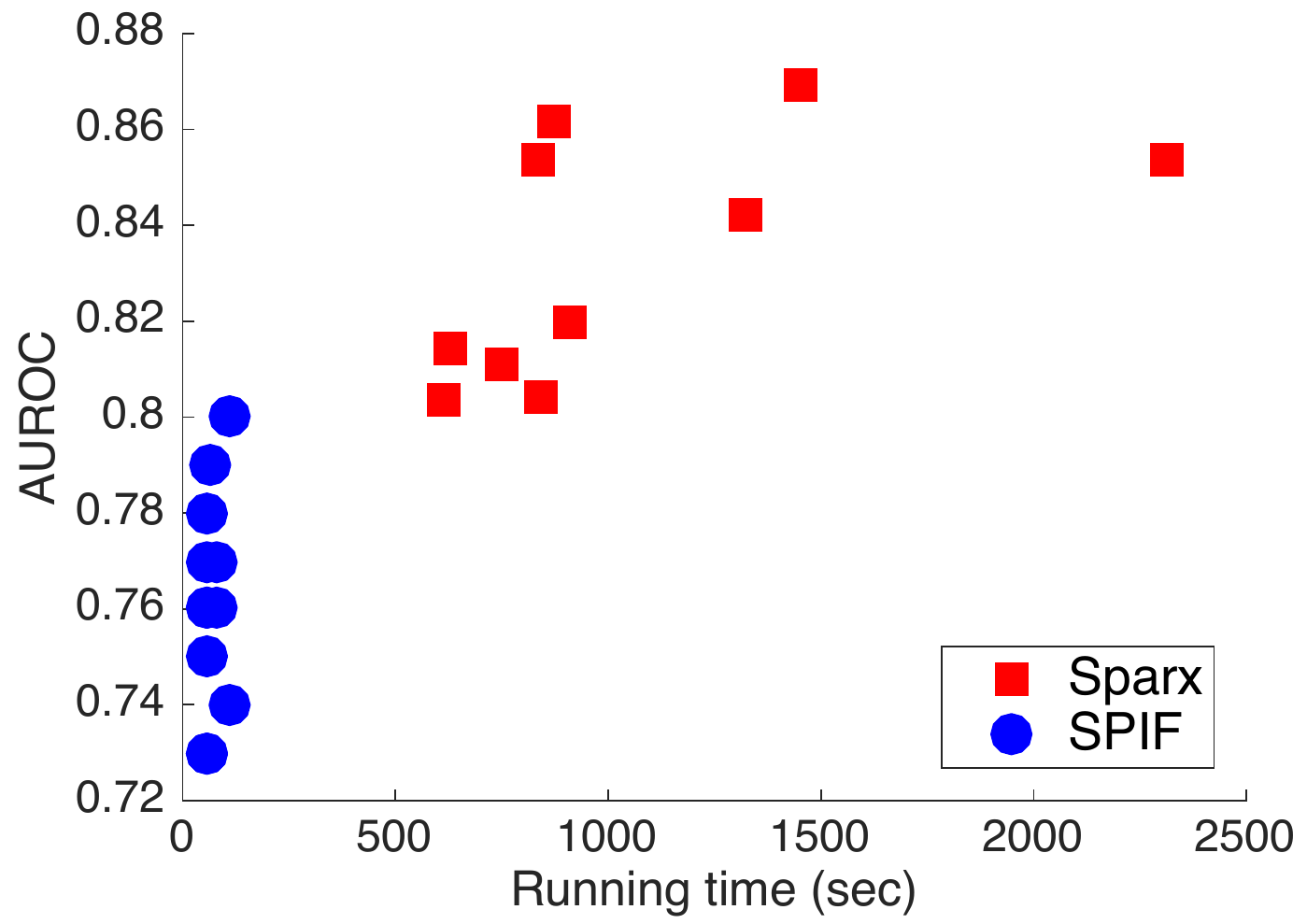}
	\hspace{0.00in}	\includegraphics[width=0.4995\linewidth]{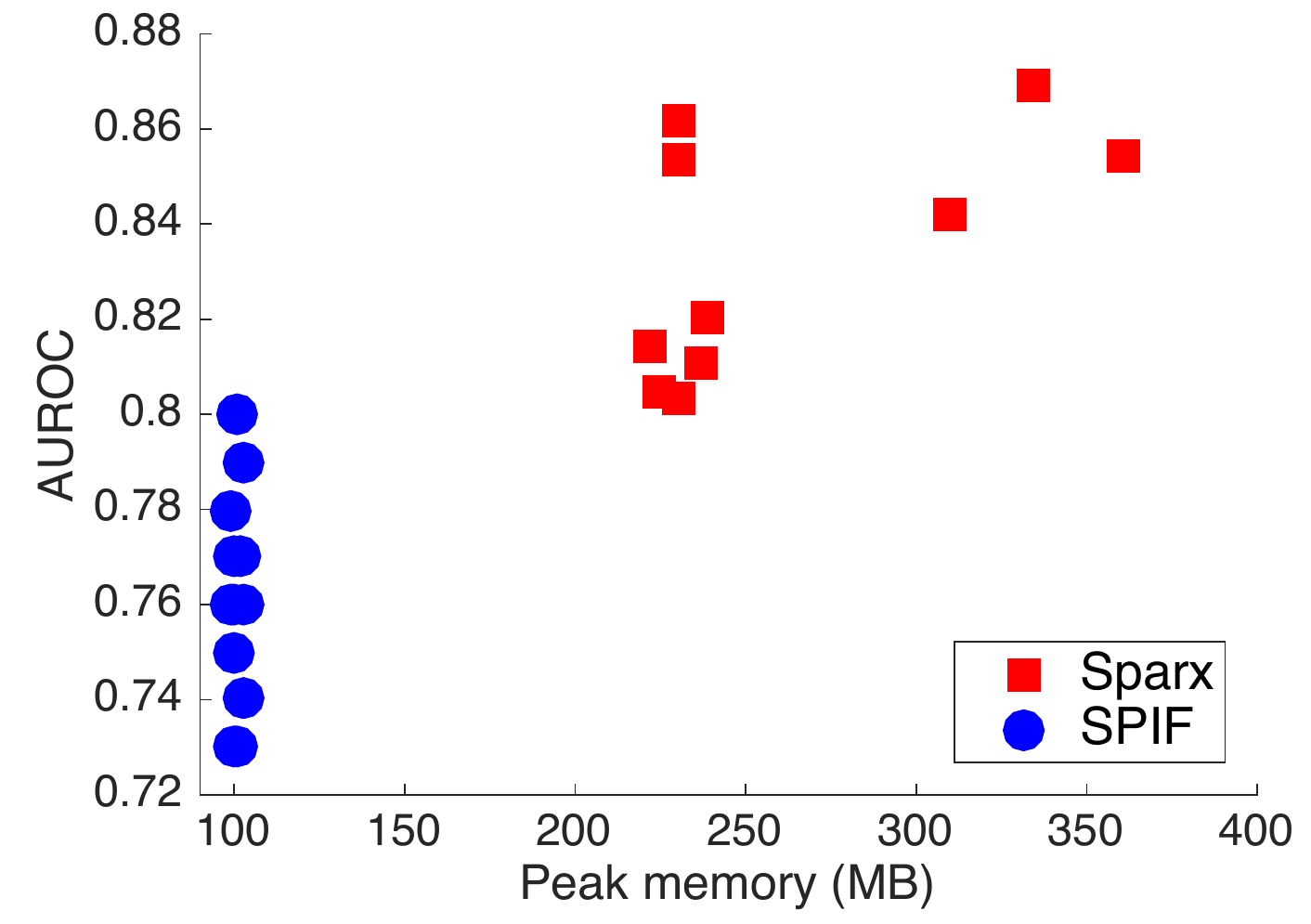}
	\vspace{-0.25in}
	\caption{Comparing \method (red) and \dif (blue) on \gis: (left) Running time (sec) vs. accuracy in AUROC, and (right) Peak driver memory (MB) vs. AUROC. Symbols depict different hyperparameter configurations. \dbs does not run on \gis due to dimensionality. \label{fig:gis}}
	\vspace{-0.15in}
\end{figure}

Next we analyze the trade-off between detection quality and resources required. 
Fig. \ref{fig:gis} shows AUROC (y-axis) versus total running time (x-axis, left) and peak driver memory (x-axis, right) under 
\congen. We find that \dif performance varies between 0.72-0.80 across HP configurations whereas \method reaches 0.80-0.87. On the other hand, \method uses more resources; as compared to \dif's typical runtime 1-2 minutes, \method can achieve its peak performance in around 14 minutes, using 2-3$\times$ more memory.
Similar conclusions are drawn under
\conmod, which is in Appx. \ref{sssec:gis}, Fig. \ref{fig:gis2}.

Table \ref{tab:h2h} presents a more ``head-to-head'' comparison under comparable settings of the HPs for both methods.
Increasing (doubling) the number of ensemble components improves \method's accuracy, with no effect on \dif; whereas it is vice versa when increasing the sampling rate. However, even with all data (sampling rate $=$ 1), \dif does not exceed an AUROC of 0.8. 

Importantly, notice that when the number of samples per tree goes up from 4,000 (rate$=$0.1) to 40,000 (rate$=$1), \dif's running time notably increases. This is due to its model-parallel yet {\bf not} data-parallel nature; where all samples per tree are shuffled (i.e. copied) over the network to a worker that is designated to construct the tree. With 100 trees, the total network communication becomes notable.
In fact, \dif quickly becomes infeasible to run on massive datasets even with a tiny subsampling rate, exactly due to this data shuffling problem, as we show in the next section.

\begin{table}[]
	\caption{``Head to head '' comparison of \method and \dif on \gis under equivalent hyperparameter configurations.}  
	\label{tab:h2h}
	\vspace{-0.15in}
	\centering
	\scalebox{0.825}{
		\begin{tabular}{lrrr|rr|rr|rr}
			\toprule
			&         &        &       &   \multicolumn{2}{c|}{AUROC} & \multicolumn{2}{c|}{Time(s)} & \multicolumn{2}{c}{Mem (MB)} \\ \hline
			conf.	& \#comp. & sampl. & depth &   Sx        &    DIF         &       Sx       &      DIF        &       Sx        &       DIF       \\
			\midrule
			1	&     \underline{50}		  &   0.01     &  10       &      0.80
			&     0.77        &             610.5 &
			58.0             &        230.2
			&      99.6        \\
			2	&    \underline{100}     &       \underline{0.01}     &  10       &     0.85      &     0.76		     &      836.0          &   58.0           &        230.7        &      99.0        \\
			3	&   100     &       \underline{0.1}     &  \underline{10}        &     0.86		        &    0.79		         &    874.2          &      61.1        &     229.9          &     102.8         \\
			4	&     100     &       \underline{0.1}     &  \underline{20}        &      0.87		       &    0.78		         &    1455.6          &          60.1    &        334.5       &       99.0       \\
			5	&    100     &       \underline{1}     &  20       &     0.85		        &    0.80	         &     2312.8         &       111.0       &         360.8      & 	101.1 \\
			\bottomrule           
		\end{tabular}
	}
\end{table}

\vspace{-0.05in}
\subsubsection{\bf \dif does not scale w.r.t. input size $n$.~} 
As a reminder, a key design principle in distributed computing on big data is ``\textit{code goes to data}'' and {\bf not} vice versa. In other words, the goal is to compute as much as possible {\em locally} and not to move around data between compute nodes (other than data containing intermediate results).
The way \dif is implemented violates this principle and suffers on large-scale data. We demonsrate the problem using \open.

Using \congen, we input to \dif\footnote{We set model HPs as num\_trees=50, max\_depth=25, and sample\_rate=0.01.} a gradually increasing fraction of \open (for \textit{fitting}, while all 2.7+ billion points are scored), starting only with around 1/1000'th of the data, as shown in Table \ref{tab:difvsn}. 
As we double the input size at every round, total running time and memory usage increase accordingly.
However, when the number of points per tree reaches around half a million, we get a system memory error and the program crashes.
As we continue to increase data size, data processing cannot reach the memory error before the 8-hour SC-budget is exhausted and we get a system time-out.

\begin{table}[H]
	\vspace{-0.1in}
	\caption{\dif does not scale up w.r.t. input size $n$.} 
	\label{tab:difvsn}
	\vspace{-0.15in}
	\centering
	\scalebox{0.85}{
		\begin{tabular}{rrrrrr}
			\toprule
			\multicolumn{1}{l}{Frac.}   & \multicolumn{1}{l}{\#pts/tree} & Time (s)                 & Mem (GB)                & AUPRC                    & AUROC                    \\
			\midrule
			\multicolumn{1}{l}{0.00128} & 35,471                          & \multicolumn{1}{r}{1396} & \multicolumn{1}{r}{454} & \multicolumn{1}{r}{0.19} & \multicolumn{1}{r}{0.987} \\
			\multicolumn{1}{l}{0.00256} & 70,943                          & \multicolumn{1}{r}{1402} & \multicolumn{1}{r}{455} & \multicolumn{1}{r}{0.27} & \multicolumn{1}{r}{0.989} \\
			0.00512                     & 141,887                         & \multicolumn{1}{r}{1531} & \multicolumn{1}{r}{461} & \multicolumn{1}{r}{0.38} & \multicolumn{1}{r}{0.991} \\
			0.01024                     & 283,774                         & \multicolumn{1}{r}{1834} & \multicolumn{1}{r}{463} & \multicolumn{1}{r}{0.42} & \multicolumn{1}{r}{0.993} \\
			0.02048                     & 567,548                         & MEM ERR                  & -                       & -                        & -                         \\
			0.04096                     & 1,135,097                        & MEM ERR                  & -                       & -                        & -                         \\
			0.08192                     & 2,270,194                        & TIMEOUT                  & -                       & -                        & -                         \\
			0.16384                     & 4,540,389                        & TIMEOUT                  & -                       & -                        & -              \\          
			\bottomrule
		\end{tabular}
	}
	\vspace{-0.15in}
\end{table}

As we have done, it is possible to fit \dif on a small subsample of a massive dataset -- to avoid this error during fitting -- and still be able to \textit{score} \textit{all} data points. However, as Table \ref{tab:difvsn} shows this incurs a sacrifice in detection performance (note esp. the AUPRC). 

\subsubsection{\bf Large-$n$/Small-$d$.~}
\begin{figure}[!t]
	\centering
	\hspace{-0.1in}	\includegraphics[width=0.4995\linewidth]{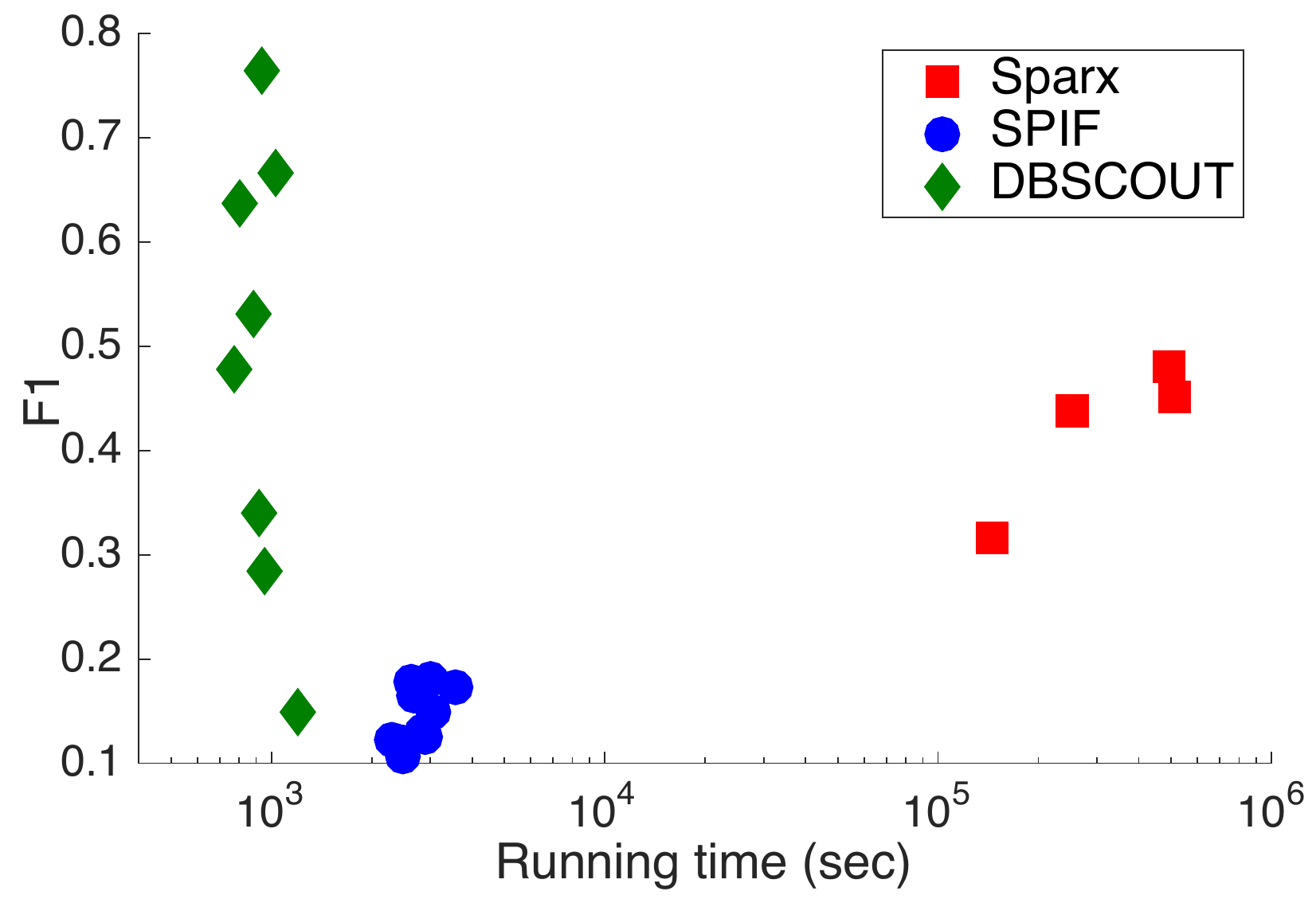}
	\hspace{0.00in}	\includegraphics[width=0.4995\linewidth]{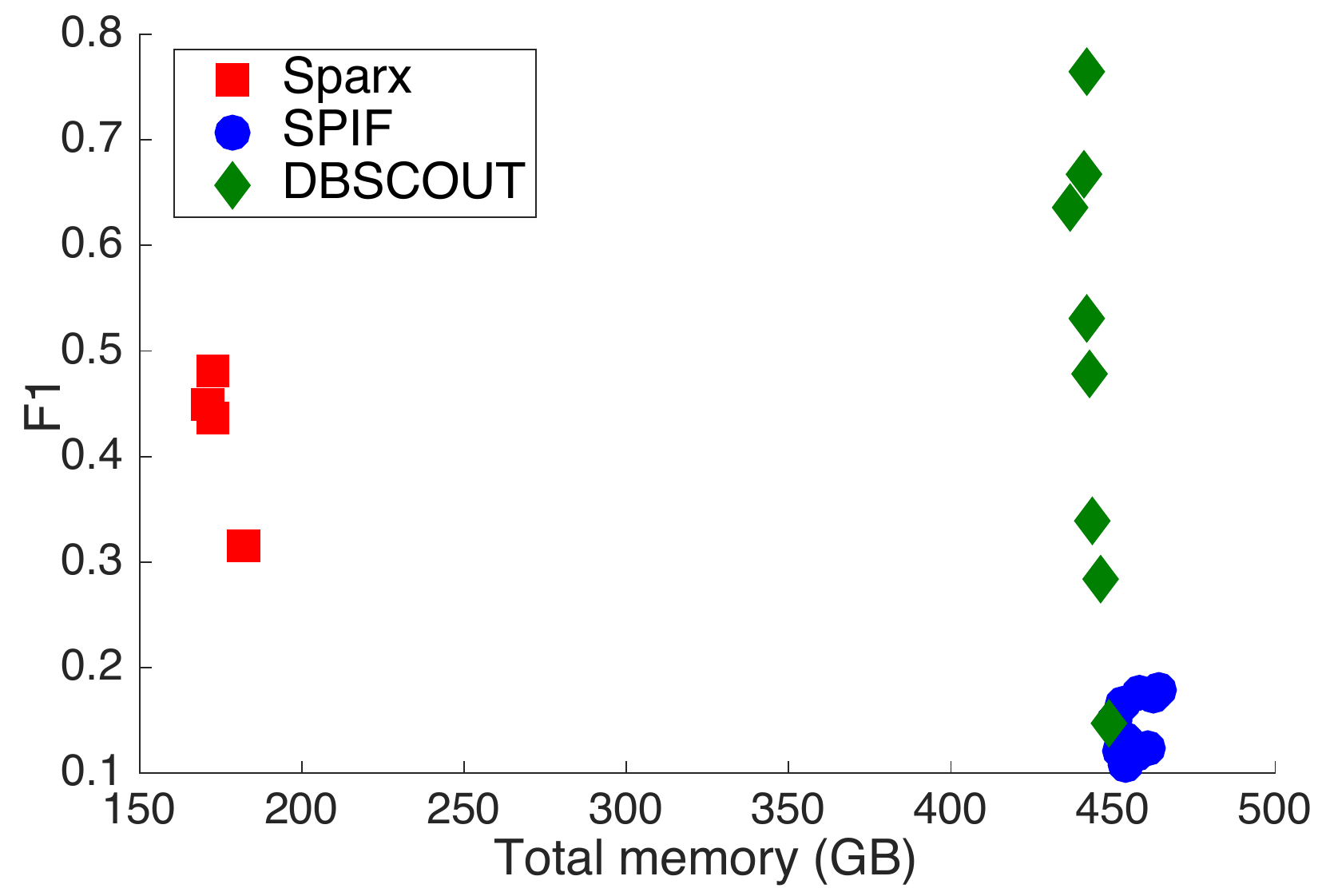}
	\vspace{-0.3in}
	\caption{Comparing all methods on \open under \congen: (left) Running time (sec) vs. accuracy in F1, and (right) Total memory (GB) vs. F1. Symbols depict different HP config.s.  \label{fig:open}}
\end{figure}

Next we analyze the results on \open, containing billions of points but only 2 dimensions.
Fig. \ref{fig:open} shows the landscape of detection performance (F1) vs. resources used for all three methods, under varying HP configurations. (See Appx. \ref{sssec:osm}, Tables \ref{table:spif-osm-3}--\ref{table:sparx-osm-3} for detailed numbers.)
Note that \dbs outputs a binary label, thus we can only report F1 for comparison.\footnote{Detailed numbers in Appx. include AUROC and AUPRC for \dif and \method.}

\dif can be fit using at most $10^{-4}$-th of the data, as discussed earlier, which results in very poor performance (F1$<$0.2).
\dbs is the fastest on this low-$d$ setting and achieves the most competitive performance, however, it is quite sensitive to the HP choices and its performance oscillates widely.
\method performance is more stable, with a longer processing time, while  using less memory.

\vspace{-0.05in}
\subsubsection{\bf Large-$n$/Large-$d$.~}
Finally, we present a similar analysis on our very large-$d$ (yet sparse) \surl.
Problematically, the
\dif implementation cannot handle sparse RDD input (and it is infeasible to store \surl as a dense RDD).
Therefore, we transform it using our random projections to $d$=100. (Our \method also uses $K$=100 projections.) 
\dbs scales very poorly with dimensionality, for which we also transform \surl to $d$=7 (largest $d$ that \dbs could handle), as well as $d$=2 (for comparison).

\begin{figure}[!ht]
	\vspace{-0.15in}
	\centering
	\hspace{-0.1in}	\includegraphics[width=0.51\linewidth]{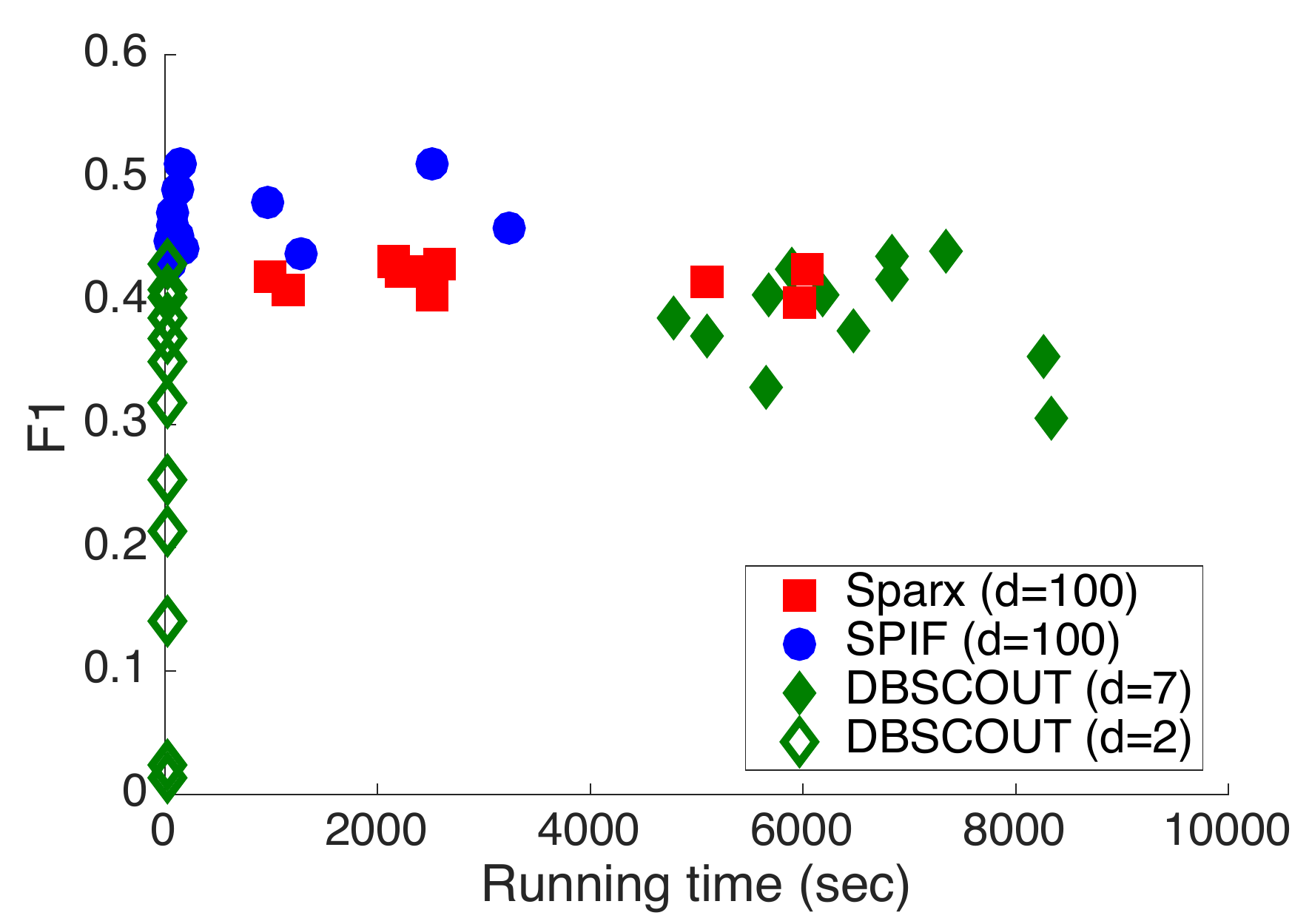}
	\hspace{-0.1in}	\includegraphics[width=0.51\linewidth]{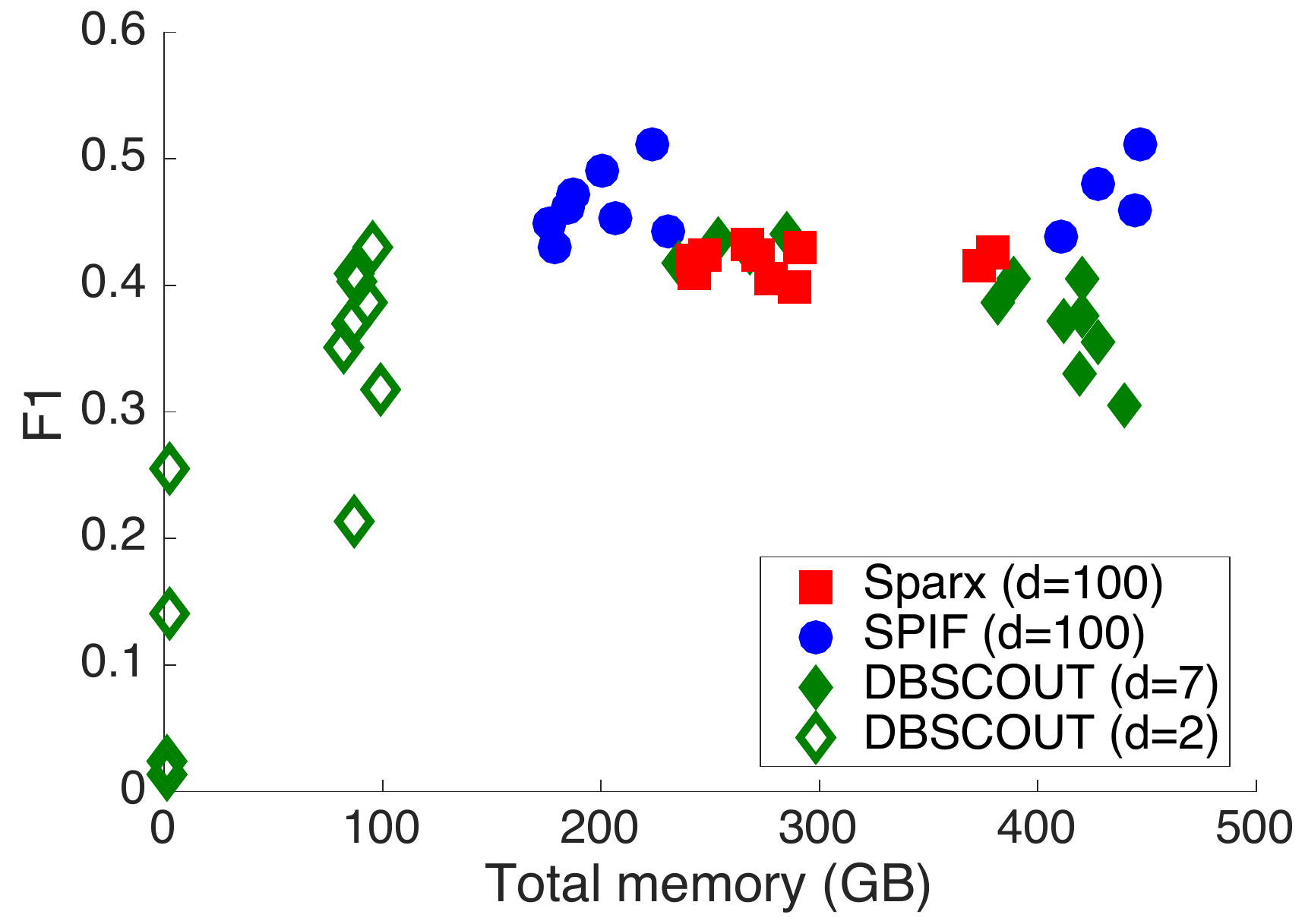}
	\vspace{-0.1in}
	\caption{Comparing methods on \surl under \congen: (left) Running time (sec) vs. accuracy in F1, and (right) Total memory (GB) vs. F1. Symbols depict different HP config.s.  \label{fig:spam}}
	\vspace{-0.1in}
\end{figure}

As shown in Fig. \ref{fig:spam}, \dbs with $d$=2 is quite resource-frugal, however it has a widely varying performance depending on the choice of hyperparameters. It achieves more stable performance with $d$=7, yet is inferior to \dif in both time and accuracy.
On the other hand, 
\method performance is robust to different hyperparameter settings, and is on par with the competing baselines.  


\vspace{-0.05in}
\subsubsection{\bf Speed-up by increasing parallelism.~} 

\begin{figure}[!th]
	\vspace{-0.1in}
	\centering
	\includegraphics[width=0.7\linewidth]{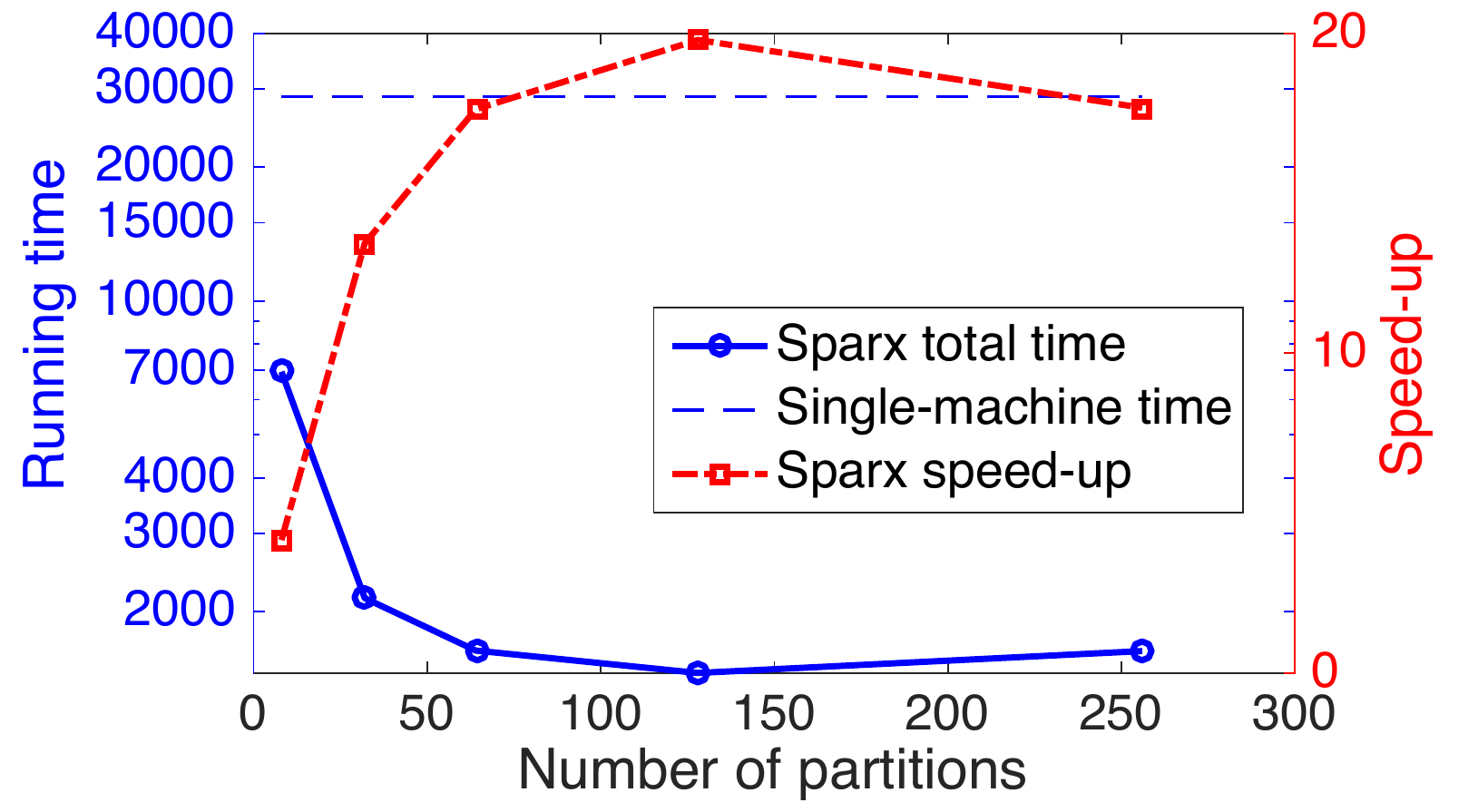}
	\vspace{-0.15in}
	\caption{(in blue) Running time of \method on \gis as number of data partitions increases. (in red) Speed-up w.r.t. single-machine implementation. \label{fig:speedup1}}
	\vspace{-0.15in}
\end{figure}

Next we show how \method leverages data-parallelism, Using \gis, we increase the number of DataFrame partitions on Spark. 
Fig. \ref{fig:speedup1} shows that  the running time decreases as partitions increase from 8 to 128, and then slightly increases for 256. This is expected behavior of distributed platforms---that speed-up is not monotonic: when the data is partitioned too much to the extent that each worker is under-utilized, the cost of network communication between workers overtakes and reduces the gains from parallelism. As compared to the running time of single-machine \ex, \method provides 4-20$\times$ speed-up.

\vspace{-0.05in}
\subsubsection{\bf \method scalability with input size $n$.~} 
\begin{figure}[!t]
	\centering
	\includegraphics[width=0.6\linewidth]{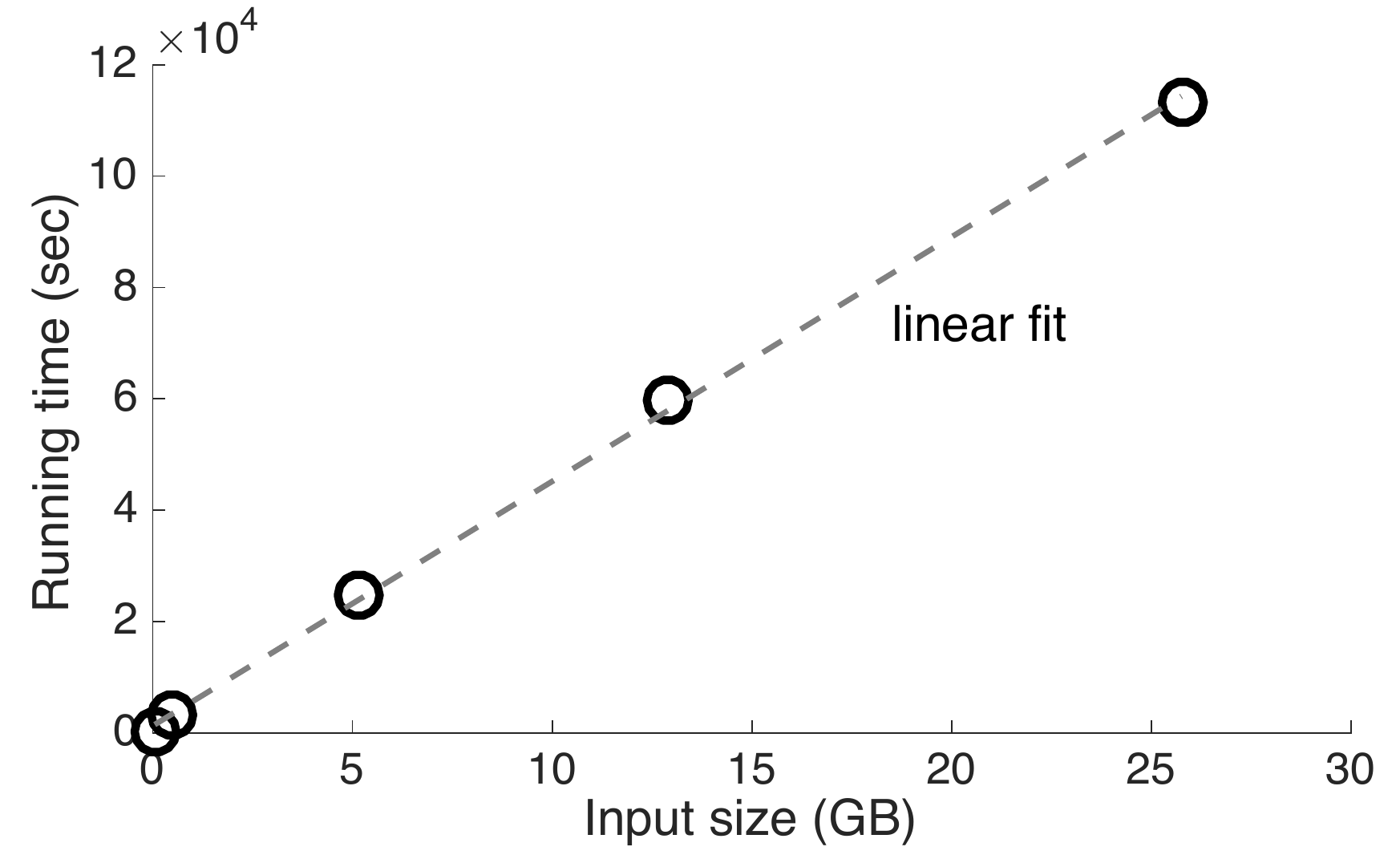}
	\vspace{-0.15in}
	\caption{\method scales linearly in number of points $n$. \label{fig:scalebyn}}
\end{figure}

Finally, we study the scalability of our distributed \method w.r.t. input size. 
Recall that dimensionality $d$ is associated with Step 1. (projection) of the algorithm, where in Sec. \ref{ssec:st} we have shown that \method is linear in $d$. We have also shown that across all 3 steps, it is linear in the number of data points $n$. Since Spark-like platforms are distributed/data-parallel in $n$, we study the running time of \method for increasing sizes of $n$ using \open.\footnote{We set model HPs as num\_chains=10, depth=5, and sample\_rate=1.} As shown in Fig. \ref{fig:scalebyn}, \method scales linearly w.r.t. $n$, emprically confirming our complexity analysis in Sec. \ref{ssec:st}.

{\bf Summary and remarks.} 
To sum up, through extensive experiments we showed that \method applies to all large-$n$ and/or large-$d$ datasets,  provides competitive detection accuracy-running time trade-off, takes advantage of data parallelism effectively, and scales up with increasing input size. In contrast, among the handful of public-domain distributed OD algorithms, (1) \dbs \textbf{has poor scalability w.r.t. dimensionality $d$}, being applicable to only small-$d$ ($d$$<$$10$) datasets in practice. On the other hand, (2) \dif \textbf{cannot handle large-$n$ datasets} due to its non-data-parallel implementation, also rendering it a non-practical choice.

\section{Related Work}
\label{sec:related}

Outlier mining has a large literature owing to its many high-stakes applications in finance, environmental monitoring, surveillance and security, to name a few. However, most existing work on point outlier detection  (OD) \cite{aggarwal2017introduction}, including those for data streams \cite{manzoor2018xstream,na2018dilof}, are designed for a single machine. Distinctly, we focus our survey on distributed detection techniques.


Although OD for large-scale data is extremely important in the big data context, and likely to become more relevant over time, there are relatively much fewer \textit{parallel} OD algorithms for truly \textit{distributed} environments with thousands of compute nodes, such as cloud services. 
A group of parallel algorithms are designed for  \textit{shared-memory} multi-core computer systems \cite{lozano2005parallel,oku2014parallel} and not for distributed settings.
Other parallel algorithms for distributed architectures are centralized; 
requiring a central ``communication/sync'' unit. 
For example, the top-$n$ OD algorithms by Angiulli {\em et al.} \cite{angiulli2012distributed,angiulli2010distributed} assume a ``supervisor'' node for synchronization. 
Similarly, Bhaduri {\em et al.} \cite{bhaduri2011algorithms}
also require a ``central'' node that maintains and updates the top-$n$ points.
Those are not applicable to modern scale-out (i.e. distributed) \textit{shared-nothing} architectures that do not employ such centralization.
Moreover, those work focus on conceptual algorithm design and do not present practical implementations.



We remark that a related category of work on distributed OD for wireless sensors \cite{palpanas2003distributed,tsou2018robust,luo2018distributed} is notably different from our work, in that those often require communication between (nearby) sensors and operate under battery/power constraints which do not apply to shared-nothing settings. 






Among distributed OD algorithms for shared-nothing architectures, Tao {\em et al.} proposed SPIF \cite{tao2018parallel}, a Spark-based design of the popular Isolation Forest algorithm \cite{liu2008isolation}.
However, SPIF emloys {\em model-parallelism} (as opposed to data-parallelism); specifically it trains each individual component (i.e. iTree) of the IF ensemble on a separate compute node. Alarmingly, the data for each iTree is shuffled to the corresponding node over the network, adding to the communication cost.  
Despite radically growing real-world datasets with billions of points, to the best of our knowledge, there are only two {\em data-parallel} (i.e., horizontally scalable \cite{corain2021dbscout}) distributed OD solutions  in the literate thus far.

Yan {\em et al.}  proposed DDLOF \cite{yan2017distributed}, the 
first distributed LOF algorithm for Hadoop MapReduce \cite{dean2008mapreduce}, and later 
its extension to top-$n$ outliers \cite{yan2017distributedtopn}. 
Besides Hadoop being orders of magnitude slower than Spark \cite{Zaharia10},  
as has been shown recently \cite{corain2021dbscout}, DDLOF fails to scale to very large
datasets.
Their grid-based data partitioning strategy makes it unsuitable for high-dimensional data as the number of partitions grows exponentially with increasing dimensionality.

Most recently, Corain {\em et al.} introduced DBSCOUT \cite{corain2021dbscout} with a public-domain
Spark implementation.
By design, it 
does not provide a ranking of the outliers (i.e. output is binary) and 
has two critical hyperparameters (\eps and \minpts) to set. 
From a scalability perspective, even though it is linear in the number of input points, it scales extremely poorly with dimensionality $d$. All of their experiments are limited to 2- or 3-$d$ datasets. 





\vspace{-0.05in}
\section{Conclusion}
\label{sec:conclusion}

We presented \method, a new scalable open-source tool for distributed outlier detection (OD).
We described its design principles and the underlying distributed/data-parallel algorithms for shared-nothing cloud-computing platforms, and open-sourced its Apache PySpark implementation  at \la{\url{https://tinyurl.com/sparx2022}}.

OD finds numerous applications, yet there are limited public-domain resources for \textit{distributed} OD as the vast majority of the literature focuses on single-machine algorithmic problems. 
Through extensive experiments, we showed that the few existing open-source tools do not match up with \method; they either do not scale well with dataset size or increasing dimensionality. Distinctly, \method is fully data-parallel, and scales linearly. 
We believe \method sets the state-of-the-art in terms of detection performance and scalability for distributed OD tasks. We expect it to increase the usability of OD on large-scale modern-day datasets that are already cloud-resident, and to offer significant impact in the applied domain for various business, engineering and scientific use cases.



\vspace{-0.05in}
{\small{
\begin{acks}
	\vspace{-0.05in}
This work is sponsored by the U.S. Army Network Enterprise Technology Command (NETCOM) and NSF CAREER 1452425. Any conclusions expressed in this material are those of the author and do not necessarily reflect the views of the funding parties.
\end{acks}
}}
\vspace{-0.05in}
\bibliographystyle{ACM-Reference-Format}
\bibliography{refs}

\clearpage

\appendix 
\section{Appendix}
\label{sec:appendix}

\subsection{Details on Experiment Setup}

\subsubsection{Ground-truth outliers in \open}
\label{appx:inject}
In previous work \cite{yan2017distributed,corain2021dbscout},  the OpenStreetMap (\open) dataset was used only for scalability and running time experiments but not for evaluation detection performance, since it does not contained labeled outliers.
In this work, we inject simulated outliers in order to evaluate and compare algorithms in all respects; detection quality, as well as resources (time and memory) used.

How we define and simulate these ground-truth outliers are as follows. 
Originally, the \open dataset contains 2,771,233,904 points of GPS coordinates, i.e. tuples of latitude and longitude values, from real-world users' travel trajectories.
To inject outliers, we first generate a 2-d histogram of the full dataset. This is done by first
creating a grid with cell size (0.01 $\times$ 0.01) that covered the full space (-180,180) $\times$ (-90,90). Then, we count the 
number of points that fall into each cell and mark all empty grid cells whose immediate 8 neighbours are also empty. 
Each outlier was then generated by randomly picking one of these marked cells and then uniformly selecting coordinates within the
cell. Our final dataset consists of 2,772,433,904  billion points, of which 1,000,000 (0.036\%) are outliers. 

The visualization of the final dataset is shown in \ref{fig:osm}, where black dots depict GPS coordinates visited by real-world users, and red dots illustrate the injected outliers.

\subsubsection{System configuration details}
\label{appx:config}
Specified in Table \ref{tab:sysconfigs}.

\begin{table}[H]
	\vspace{-0.15in}
	\caption{Two different system config.s used in experiments; `moderate' \conmod and more `generous' \congen.}  
	\label{tab:sysconfigs}
	\vspace{-0.15in}
	\centering
	\scalebox{0.85}{
		\begin{tabular}{lcccccc}
			\toprule
			&  \#partitions & driver  & exec & \#execs & \#exec  & \#threads \\
			&				& memory & memory &		& cores & \\
			\midrule
			\conmod	& 64 & 25GB & 4GB & 4 & 4 &  4\\
			\congen	& 128 & 45GB  & 8GB & 64 & 8 & 128 \\
			\bottomrule
			\vspace{-0.3in}
		\end{tabular}
	}
\end{table}

\subsection{Additional/Detailed Experiment Results}
\label{appx:additional}

\subsubsection{\bf Results on \gis under \conmod~} 
\label{sssec:gis} 

\begin{figure}[H]
	\vspace{-0.15in}
	\centering
	\hspace{-0.1in}	\includegraphics[width=0.495\linewidth]{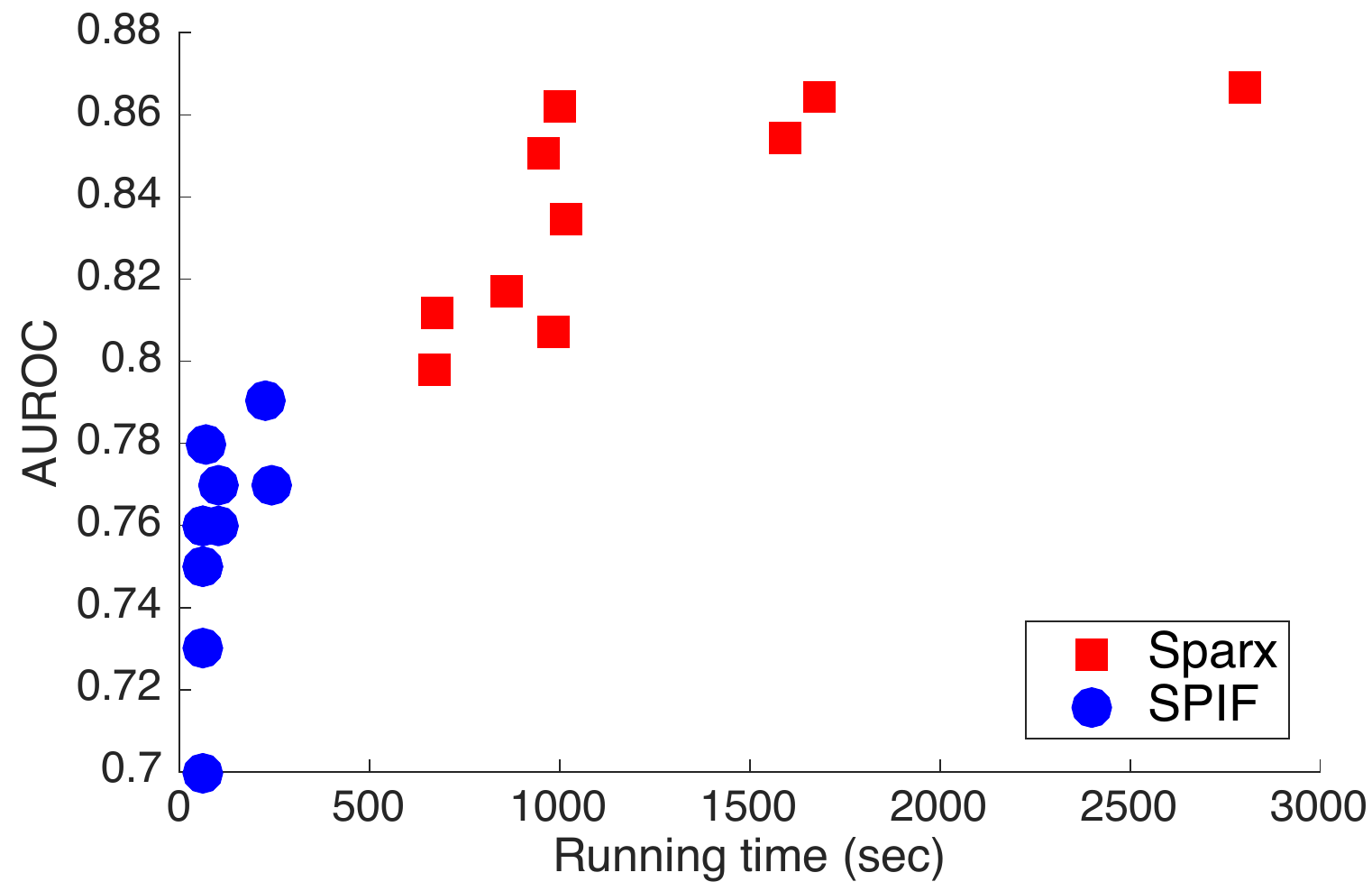}
	\hspace{0.00in}	\includegraphics[width=0.495\linewidth]{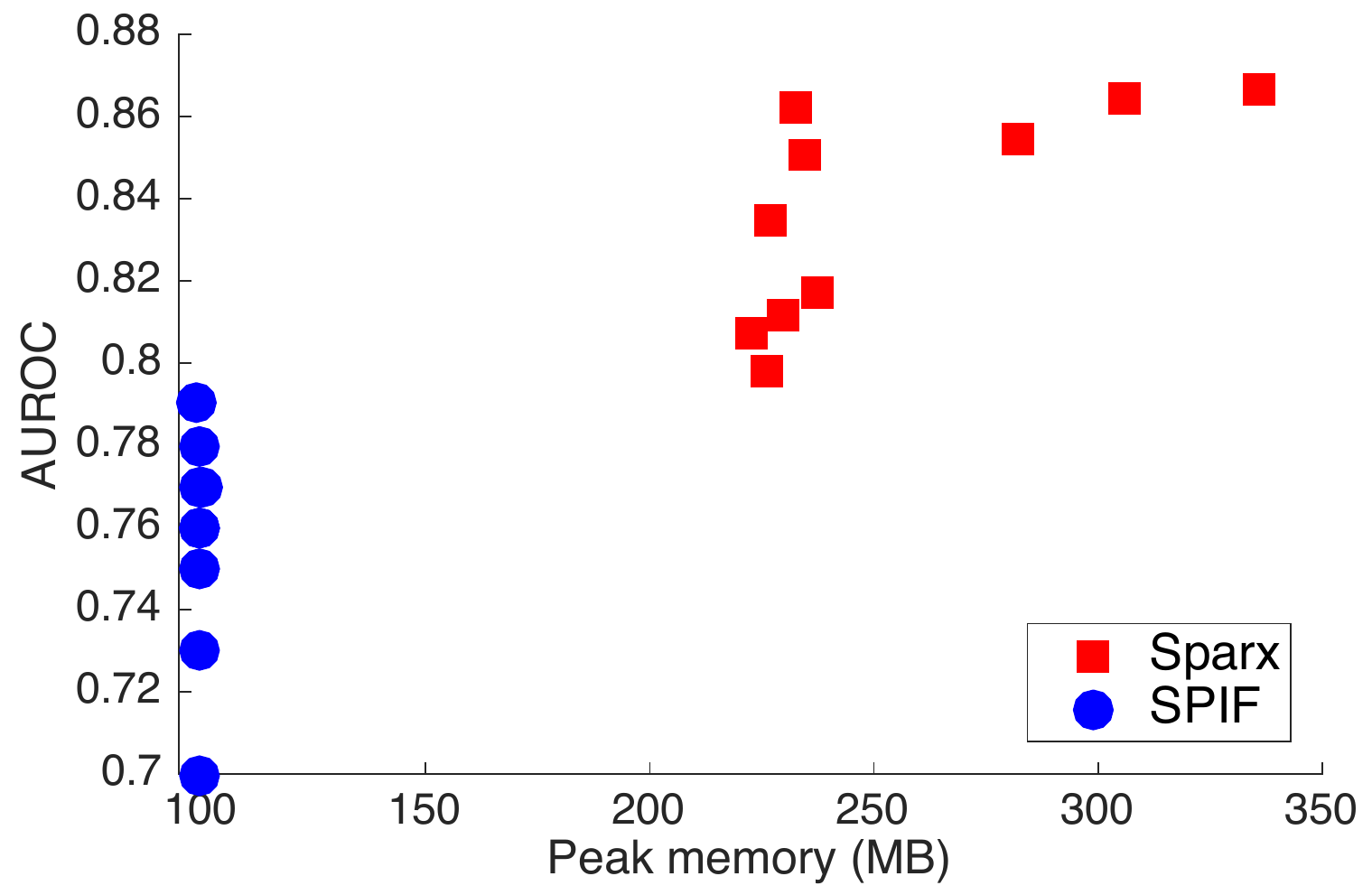}
	\vspace{-0.15in}
	\caption{Using \conmod: (left) Running time (sec) vs. accuracy in AUROC, and (right) Peak driver memory (MB) vs. AUROC on \gis comparing \method (red) and \dif (blue). \dbs does not run on \gis due to dimensionality. \label{fig:gis2}}
	\vspace{-0.1in}
\end{figure}

In Fig. \ref{fig:gis2} We showcase the AUROC performance versus running time (left) and peak memory (right) on \gis under the moderate system configuration \conmod.
Results are for \method and \dif only, since \dbs cannot scale beyond about 10 dimensions.

\begin{table}[t]
	\setlength{\tabcolsep}{1pt}
	\caption{\dif performance and resources used on \open under \conmod and varying HP configurations. Note that fewer config.s can be handled under moderate resources as compared to \congen.
		Sampling rate is from 1\% of the original data, and thus should be multiplied further by 0.01$\times$.
		\label{table:spif-osm-1}}
	\vspace{-0.15in}
	\scalebox{0.85}{
		\begin{tabular}{rcc|cc|ccc}
			\toprule
			\multicolumn{1}{c}{\textbf{\#comp.}} & \multicolumn{1}{c}{\textbf{depth}} & \multicolumn{1}{c|}{\textbf{sampl.}} & \multicolumn{1}{c}{\textbf{Time(s)}} & \multicolumn{1}{c|}{\textbf{Mem(GB)}} & \multicolumn{1}{c}{\textbf{AUROC}} & \multicolumn{1}{c}{\textbf{AUPRC}} & \multicolumn{1}{c}{\textbf{F1}} \\
			\midrule
			50                         & 10                        & 0.001                      & 2284                 & 478                  & 0.978                & 0.161                & 0.115                \\
			50                         & 10                        & 0.005                      & 2562                 & 481                  & 0.980                 & 0.327                & 0.164                \\
			50                         & 20                        & 0.001                      & 2530                 & 481                  & 0.986                & 0.149                & 0.118                \\
			50                         & 20                        & 0.005                      & 2908                 & 486                  & 0.991                & 0.389                & 0.162                \\
			100                        & 10                        & 0.001                      & 2920                 & 480                  & 0.979                & 0.144                & 0.118                \\
			100                        & 20                        & 0.001                      & 3475                 & 483                  & 0.986                & 0.155                & 0.123    \\
			\bottomrule           
		\end{tabular}
	}
\end{table}

\subsubsection{\bf Detailed results on \open~} 
\label{sssec:osm}

Tables \ref{table:spif-osm-1} and \ref{table:spif-osm-3} (under \conmod and \congen, respectively) provide performance and resource usage details of \dif on \open with varying hyperparameter settings.
Notice that we subsample \open substantially for being able to fit \dif, as not being data-parallel, it does not scale well with $n$.
\begin{table}[H]
	\setlength{\tabcolsep}{1pt}
	\caption{\dif performance and resources used on \open under \congen and varying hyperparameter (HP) configurations.
		Sampling rate is from 1\% of the original data, and thus should be multiplied further by 0.01$\times$.
		\label{table:spif-osm-3}}
	\vspace{-0.15in}
	\scalebox{0.85}{
		\begin{tabular}{rcc|cc|ccc}
			\toprule
			\multicolumn{1}{c}{\textbf{\#comp.}} & \multicolumn{1}{c}{\textbf{depth}} & \multicolumn{1}{c|}{\textbf{sampl.}} & \multicolumn{1}{c}{\textbf{Time(s)}} & \multicolumn{1}{c|}{\textbf{Mem(GB)}} & \multicolumn{1}{c}{\textbf{AUROC}} & \multicolumn{1}{c}{\textbf{AUPRC}} & \multicolumn{1}{c}{\textbf{F1}} \\
			\midrule
			50                                   & 10                                 & 0.001                               & 2292                                 & 452                                  & 0.978                              & 0.139                              & 0.122                           \\ 
			50                                   & 10                                 & 0.005                               & 2662                                 & 453                                  & 0.980                              & 0.311                              & 0.165                           \\
			50                                   & 10                                 & 0.010                               & 2474                                 & 457                                  & 0.978                              & 0.346                              & 0.119                           \\
			50                                   & 20                                 & 0.001                               & 2463                                 & 454                                  & 0.986                              & 0.150                              & 0.107                           \\
			50                                   & 20                                 & 0.005                               & 2627                                 & 458                                  & 0.991                              & 0.367                              & 0.177                           \\
			50                                   & 20                                 & 0.010                               & 2975                                 & 464                                  & 0.992                              & 0.439                              & 0.180                           \\
			100                                  & 10                                 & 0.001                               & 2844                                 & 454                                  & 0.978                              & 0.166                              & 0.131                           \\
			100                                  & 10                                 & 0.005                               & 3076                                 & 450                                  & 0.979                              & 0.324                              & 0.150                           \\
			100                                  & 20                                 & 0.001                               & 2913                                 & 461                                  & 0.986                              & 0.159                              & 0.124                           \\
			100                                  & 20                                 & 0.005                               & 3555                                 & 462                                  & 0.991                              & 0.387                              & 0.174      \\            
			\bottomrule         
		\end{tabular}
	}
	\vspace{-0.05in}
\end{table}

Tables \ref{table:dbs-osm-3} and \ref{table:dbs-osm-1} (under \congen and \conmod, respectively) provide performance and resource usage details of \dbs on \open with varying hyperparameter settings.
\dbs excels on this very low dimensional dataset, as it is designed accordingly.

\begin{table}[H]
	\vspace{-0.1in}
	\caption{\dbs performance and resources used on \open under \congen and varying HP configurations. 
		Note that 
		\dbs output is binary and thus only F1 is reported. \label{table:dbs-osm-3}}
	\vspace{-0.15in}
	\scalebox{0.85}{
		\begin{tabular}{rr|rc|c}
			\toprule
			\multicolumn{1}{r}{\textbf{minPts}} & \multicolumn{1}{r|}{\textbf{eps.}} & \multicolumn{1}{r}{\textbf{Time(s)}} & \multicolumn{1}{r|}{\textbf{Mem(GB)}} & \multicolumn{1}{r}{\textbf{F1}} \\
			\midrule
			100                                 & 250000                            & 957                                  & 446                                  & 0.283                           \\
			100                                 & 500000                            & 775                                  & 443                                  & 0.478                           \\
			100                                 & 1000000                           & 799                                  & 437                                  & 0.637                           \\
			100                                 & 2000000                           & 938                                  & 442                                  & 0.765                           \\
			200                                 & 250000                            & 1209                                 & 449                                  & 0.148                           \\
			200                                 & 500000                            & 918                                  & 444                                  & 0.339                           \\
			200                                 & 1000000                           & 884                                  & 442                                  & 0.531                           \\
			200                                 & 2000000                           & 1030                                 & 441                                  & 0.667      \\
			\bottomrule                    
		\end{tabular}
	}
	\vspace{-0.25in}
\end{table}
\begin{table}[H]
	\caption{\dbs performance and resources used on \open under \conmod and varying hyperparameter configurations. Note that \dbs output is binary and thus only F1 is reported.
		\dbs scales well to this large-$n$/small-$d$ (2-d) dataset, w/ comparable results to those under \congen.
		\label{table:dbs-osm-1}}
	\vspace{-0.15in}
	\scalebox{0.85}{
		\begin{tabular}{rr|rc|c}
			\toprule
			\multicolumn{1}{r}{\textbf{minPts}} & \multicolumn{1}{r|}{\textbf{eps.}} & \multicolumn{1}{r}{\textbf{Time(s)}} & \multicolumn{1}{r|}{\textbf{Mem(GB)}} & \multicolumn{1}{r}{\textbf{F1}} \\
			\midrule
			100                                 & 250000                            & 1279                                 & 474                                  & 0.283                           \\
			100                                 & 500000                            & 911                                  & 469                                  & 0.478                           \\
			100                                 & 1000000                           & 855                                  & 468                                  & 0.637                           \\
			100                                 & 2000000                           & 1167                                 & 466                                  & 0.764                           \\
			200                                 & 250000                            & 1615                                 & 478                                  & 0.148                           \\
			200                                 & 500000                            & 1031                                 & 468                                  & 0.339                           \\
			200                                 & 1000000                           & 930                                  & 466                                  & 0.531                           \\
			200                                 & 2000000                           & 1257                                 & 468                                  & 0.667                        \\
			\bottomrule  
		\end{tabular}
	}
	\vspace{-0.1in}
\end{table}
\begin{table}[H]
	\setlength{\tabcolsep}{1pt}
	\caption{\method performance and resources used on \open under \congen and varying hyperparameter configurations (sampling rate is set to 0.01). \label{table:sparx-osm-3}}
	\vspace{-0.15in}
	\scalebox{0.95}{
		\begin{tabular}{rc|cc|ccc}
			\toprule
			\multicolumn{1}{c}{\textbf{\#comp.}} & \multicolumn{1}{c|}{\textbf{depth}} &  \multicolumn{1}{c}{\textbf{Time(s)}} & \multicolumn{1}{c|}{\textbf{Mem(GB)}} & \multicolumn{1}{c}{\textbf{AUROC}} & \multicolumn{1}{c}{\textbf{AUPRC}} & \multicolumn{1}{c}{\textbf{F1}} \\
			\midrule
			10                          & 5                         & 144041                      & 182.36                      & 0.959                     & 0.271                     & 0.316                  \\
			10                          & 10                        & 254243                      & 172.89                      & 0.973                     & 0.400                     & 0.437                  \\
			20                          & 10                        & 509014                      & 170.72                      & 0.974                     & 0.443                     & 0.451                  \\
			10                          & 20                        &492506        & 172.85        & 0.975     & 0.446      & 0.480
			\\
			\bottomrule  
		\end{tabular}
	}
\end{table}

\subsubsection{\bf Detailed results on \surl~}  Table \ref{table:spif-spam100-1} provides performance and resource usage details of \dif on \surl with varying hyperparameter settings under \conmod.

Note that \surl is very high dimensional yet sparse, however the
\dif implementation cannot handle sparse RDD input (and it is infeasible to store \surl as a dense RDD).
Therefore, we transform it using our random projections to $d$=100. (Our \method also uses $K$=100 projections.) 

\begin{table}[H]
	\setlength{\tabcolsep}{1pt}
	\caption{\dif performance and resources used on \surl ($d$$=$100) under \conmod and varying hyperparameter configurations.
		The	\textbf{best} (in bold) and the \underline{worst} F1 performance highlighted (only measure \dbs can be compared to). 
		\label{table:spif-spam100-1}}
	\vspace{-0.15in}
	\scalebox{0.95}{
		\begin{tabular}{rrr|rr|ccc}
			\toprule
			\multicolumn{1}{c}{\textbf{\#comp.}} & \multicolumn{1}{c}{\textbf{depth}} & \multicolumn{1}{c|}{\textbf{sampl.}} & \multicolumn{1}{c}{\textbf{Time(s)}} & \multicolumn{1}{c|}{\textbf{Mem(GB)}} & \multicolumn{1}{c}{\textbf{AUROC}} & \multicolumn{1}{c}{\textbf{AUPRC}} & \multicolumn{1}{c}{\textbf{F1}} \\
			\midrule
			50                   & 10                   & 0.01                 & 61.8                 & 206                  & 0.656                & 0.479                & 0.463                \\
			50                   & 10                   & 0.1                  & 136.0                & 279                  & 0.703                & 0.524                & \textbf{0.526}                \\
			50                   & 20                   & 0.01                 & 63.0                 & 201                  & 0.684                & 0.502                & 0.488                \\
			50                   & 20                   & 0.1                  & 128.8                & 285                  & 0.677                & 0.491                & 0.468                \\
			100                  & 10                   & 0.01                 & 80.6                 & 202                  & 0.689                & 0.503                & 0.498                \\
			100                  & 10                   & 0.1                  & 150.8                & 318                  & 0.676                & 0.484                & 0.475                \\
			100                  & 20                   & 0.01                 & 83.0                 & 204                  & 0.639                & 0.457                & \underline{0.434}                \\
			100                  & 20                   & 0.1                  & 159.5                & 328                  & 0.659                & 0.475                & 0.451                \\
			50                   & 10                   & 1                    & 938.8                & 451                  & 0.675                & 0.492                & 0.481                \\
			50                   & 20                   & 1                    & 1192.5               & 440                  & 0.637                & 0.461                & 0.439     \\
			100                   & 10                   & 1                    & 2520.1               & 440                  & 0.637                & 0.461                & 0.439              
			\\
			\bottomrule                  
		\end{tabular}
	}
\end{table}

Tables \ref{table:dbs-spam7-1} and \ref{table:dbs-spam2-1} (under $d$$=$7 and $d$$=$2, respectively) provide performance and resource usage details of \dbs on \surl with varying hyperparameter settings under \conmod.
(As a simple heuristic, we set \minpts to 2$\times d$ and then carefully choose \eps via the elbow-method as explained in \cite{corain2021dbscout}.)

As a reminder, \dbs scales very poorly with dimensionality and cannot handle the original \surl dataset.
Therefore,
as with \dif, we reduce dimensionality via random projections. The largest $d$ that \dbs was able to handle is $d$=7, and we also report results with $d$=2 for comparison. 
Resources (time and memory) reduce for the latter, however at the expense of detection performance.

Table \ref{table:sparx-spam100-3} provides performance and resource usage details of \method on \surl with varying hyperparameter settings under \conmod.
Note that its performance is quite stable/robust to varying hyperparameter choices.

\begin{table}[H]
	\caption{\dbs performance and resources used on \surl ($d$$=$7) under \conmod and varying hyperparameter configurations.
		The	\textbf{best} (in bold) and the \underline{worst} performance highlighted. \label{table:dbs-spam7-1}}
	\vspace{-0.15in}
	\scalebox{0.95}{
		\begin{tabular}{rr|rr|r}
			\toprule
			\multicolumn{1}{r}{\textbf{minPts}} & \multicolumn{1}{r|}{\textbf{eps.}} & \multicolumn{1}{r}{\textbf{Time(s)}} & \multicolumn{1}{r|}{\textbf{Mem(GB)}} & \multicolumn{1}{r}{\textbf{F1}} \\
			\midrule
			14                         & 0.6                      & 9589                        & 233                         & 0.418                  \\
			14                         & 0.7                      & 8860                        & 276                         & 0.426                  \\
			14                         & 0.8                      & 7739                        & 385                         & 0.386                  \\
			14                         & 0.9                      & 12424                       & 419                         & 0.371                  \\
			14                         & 0.95                     & 5697                        & 430                         & 0.330                  \\
			14                         & 1                        & 15616                       & 438                         & \underline{0.306}                  \\
			28                         & 0.6                      & 10272                       & 235                         & 0.437                  \\
			28                         & 0.7                      & 8591                        & 284                         & \textbf{0.441}                  \\
			28                         & 0.8                      & 9269                        & 382                         & 0.406                  \\
			28                         & 0.9                      & 12000                       & 421                         & 0.405                  \\
			28                         & 1                        & 8221                        & 432                         & 0.356 	\\
			\bottomrule                  
		\end{tabular}
	}
\end{table}

\begin{table}[H]
	\caption{\dbs  performance and resources used on \surl ($d$$=$2) under \conmod and varying hyperparameter configurations.
		The	\textbf{best} (in bold) and the \underline{worst} performance highlighted. 
		Notice that with only $d$$=$2, \dbs performs notably worse than that for $d$$=$7, while in turn, correspondingly lower resources are required.
		\label{table:dbs-spam2-1}}
	\vspace{-0.15in}
	\scalebox{0.95}{
		\begin{tabular}{rr|rr|r}
			\toprule
			\multicolumn{1}{r}{\textbf{minPts}} & \multicolumn{1}{r|}{\textbf{eps.}} & \multicolumn{1}{r}{\textbf{Time(s)}} & \multicolumn{1}{r|}{\textbf{Mem(GB)}} & \multicolumn{1}{r}{\textbf{F1}} \\
			\midrule
			4                          & 0.0001                   & 476                         & 116                         & 0.410                   \\
			4                          & 0.0005                   & 429                         & 130                         & 0.370                  \\
			4                          & 0.001                    & 559                         & 2.2                         & 0.352                  \\
			4                          & 0.005                    & 1139                        & 1.83                        & 0.256                  \\
			4                          & 0.01                     & 1158                        & 1.97                        & 0.141                  \\
			4                          & 0.05                     & 538                         & 1.37                        & \underline{0.013}                 \\
			8                          & 0.0001                   & 854                         & 126                         & \textbf{0.431}                  \\
			8                          & 0.0005                   & 544                         & 118                         & 0.403                  \\
			8                          & 0.001                    & 1139                        & 129                         & 0.386                  \\
			8                          & 0.005                    & 1129                        & 129                         & 0.318                  \\
			8                          & 0.01                     & 817                         & 1.27                        & 0.213                  \\
			8                          & 0.05                     & 201                         & 1.7                         & 0.023                 	\\
			\bottomrule                  
		\end{tabular}
	}
\end{table}

\begin{table}[H]
	\setlength{\tabcolsep}{1pt}
	\caption{\method performance and resources used on \surl ($K$$=$100) under \conmod and varying hyperparameter configurations.
		The	\textbf{best} (in bold) and the \underline{worst} F1 performance highlighted (only measure \dbs can be compared to). 
		\label{table:sparx-spam100-3}}
	\vspace{-0.15in}
	\scalebox{0.95}{
		\begin{tabular}{rrr|rr|ccc}
			\toprule
			\multicolumn{1}{c}{\textbf{\#comp.}} & \multicolumn{1}{c}{\textbf{depth}} & \multicolumn{1}{c|}{\textbf{sampl.}} & \multicolumn{1}{c}{\textbf{Time(s)}} & \multicolumn{1}{c|}{\textbf{Mem(GB)}} & \multicolumn{1}{c}{\textbf{AUROC}} & \multicolumn{1}{c}{\textbf{AUPRC}} & \multicolumn{1}{c}{\textbf{F1}} \\
			\midrule
			50  & 10 & 0.01 & 980.5  & 241 & 0.602 & 0.419 & 0.420 \\
			50  & 10 & 0.1  & 1150.1 & 243 & 0.590 & 0.407 & 0.410 \\
			50  & 20 & 0.01 & 2160.2 & 267 & 0.620 & 0.43  & \textbf{0.433} \\
			50  & 20 & 0.1  & 2523.0 & 277 & 0.595 & 0.409 & 0.406  \\
			100 & 10 & 0.01 & 2324.1 & 271 & 0.613 & 0.424 & 0.423  \\
			100 & 10 & 0.1  & 2584.8 & 290 & 0.617  & 0.429 & 0.430 \\
			100 & 20 & 0.01 & 5089.6 & 372 & 0.600 & 0.419 & 0.416 \\
			100 & 20 & 0.1  & 6036.6 & 379  & 0.614 & 0.428  & 0.426 \\
			50  & 10 & 1    & 2223.5  & 247  & 0.609 & 0.421 & 0.424 \\
			50  & 20 & 1    & 5965.4 & 288 & 0.577 & 0.403 & \underline{0.399} \\
			\bottomrule
		\end{tabular}
	}
\end{table}

\end{document}